# A Wavelet-Based Framework for Mapping Long Memory in Resting-State fMRI: Age-Related Changes in the Hippocampus from the ADHD-200 Dataset

**Yasaman Shahhosseini**[1] | **Cédric Beaulac**[2] | **Farouk S. Nathoo**[1] | **Michelle F. Miranda**[1]

[1]Department of Mathematics and Statistics, University of Victoria, Victoria, BC, Canada

[2]Department of Mathematics, Université du Québec à Montréal, Montréal, QC, Canada

**Correspondence**
Michelle F. Miranda
Email: michellemiranda@uvic.ca

**Abstract**

Functional magnetic resonance imaging (fMRI) time series are known to exhibit long-range temporal dependencies that challenge traditional modeling approaches. In this study, we propose a novel computational pipeline to characterize and interpret these dependencies using a long-memory (LM) framework, which captures the slow, power-law decay of autocorrelation in resting-state fMRI (rs-fMRI) signals. The pipeline involves voxelwise estimation of LM parameters via a wavelet-based Bayesian method, yielding spatial maps that reflect temporal dependence across the brain. These maps are then projected onto a lower-dimensional space via a composite basis and are then related to individual-level covariates through group-level regression. We applied this framework to the ADHD-200 dataset. Analysis identified a significant positive association between age and the long-memory parameter in the hippocampus after adjusting for ADHD symptom severity and medication status. After accounting for acquisition site, the estimated effect sizes were attenuated and no longer reached statistical significance, although the spatial pattern was largely preserved, with the strongest estimated effects remaining localized to the same anatomical regions. These findings suggest a relationship between long-range temporal dependence and developmental changes in memory-related brain regions while highlighting the influence of acquisition-site variability on effect magnitude and statistical significance. Overall, the proposed methodology enables detailed mapping of intrinsic temporal dynamics in rs-fMRI and offers new insights into the relationship between functional signal memory and brain development.

## 1 | INTRODUCTION

Functional Magnetic Resonance Imaging (fMRI) is a non-invasive neuroimaging technique that measures brain activity by detecting changes in blood oxygenation levels over time. This relies on the Blood Oxygen Level Dependent (BOLD) signal (Ogawa et al. 1990), which reflects local changes in the balance between oxygenated and deoxygenated hemoglobin following neural activation, providing an indirect measure of brain activity. Resting-state fMRI (rs-fMRI) enables the study of intrinsic functional networks without task-based stimuli and is widely used in both basic and clinical neuroscience (Bijsterbosch et al. 2017). In this study, we propose a novel voxelwise Bayesian framework to estimate long-memory properties of rs-fMRI time series and examine their associations with individual-level clinical covariates. We demonstrate the approach using the ADHD-200 preprocessed dataset (Bellec et al. 2017), which includes resting-state scans and behavioral measures from children and adolescents.

While most resting-state fMRI studies focus on spatial patterns of functional connectivity, less attention has been given to the temporal structure of the BOLD signal itself. In this work, we investigate whether the long-memory (LM) parameter offers

**Abbreviations:** ADHD, Attention-Deficit/Hyperactivity Disorder; LM, Long-Memory, fMRI, functional Magnetic Resonance Imaging.

a meaningful way to characterize intrinsic brain activity and relate it to covariates of interest. Accurately modeling temporal dependencies is essential for analyzing rs-fMRI data. While traditional autoregressive (AR) models have been widely used Woolrich et al. (2001), Lenoski et al. (2008), they assume short-range memory and may not fully capture the rich temporal organization of the BOLD signal, which has been shown to exhibit dependencies extending several minutes into the past Tagliazucchi et al. (2013). Our framework explicitly models these long-range dependencies using a long-memory process. Previous studies Zarahn et al. (1997), Aguirre et al. (1997) accounted for long-range effects through correlated noise in regression models, whereas we model each voxels signal as a long-memory process and use spatial regression to relate these estimates to subject-level covariates Sokunbi et al. (2014). This approach allows us to explore complex temporal dynamics that may reflect underlying neural processes beyond what is captured by traditional connectivity-based analyses.

The nonlinear and scale-free nature of fMRI signals has been explored in numerous studies Bullmore et al. (2004), Vazquez and Noll (1998), Birn et al. (2001), Bullmore et al. (2001), Xie et al. (2008). Although long-memory models are not explicitly nonlinear, they capture temporal complexity that reflects persistent, scale-invariant patterns in the BOLD signalfeatures that are often associated with nonlinear or fractal behavior in neuroimaging data. In our framework, we model fMRI time series at each voxel as long-memory processes, which results in dense covariance structures. To address the resulting computational burden, we apply the discrete wavelet transform (DWT) (Mallat 1989), which transforms the data from the time domain to the time-frequency domain, yielding a near-diagonal covariance structure with power-law decay Fadili and Bullmore (2002), Meyer (2003), Jeong et al. (2013), Maxim et al. (2005). This transformation facilitates efficient estimation of long-memory parameters and enables the analysis of brain signal dynamics across multiple temporal scales.

A variety of methods have been proposed to analyze the dynamics of fMRI time series. These include connectivity-based approaches such as time-resolved graph theory Maffei and Sessa (2021) and dynamic functional connectivity measures Pedersen et al. (2017), which examine how functional connections between brain regions fluctuate over time. Entropy-based methods evaluate the temporal complexity of fMRI signals by quantifying the degree of signal unpredictability or irregularity, and are often used to assess neural efficiency, cognitive state, or disease-related disruptions Omidvarnia et al. (2022), McDonough and Nashiro (2014), Nezafati et al. (2020). In contrast, our work adopts a long-memory modeling approach as a distinct spatiotemporal framework for examining brain temporal complexity through power-law scaling behavior La Rocca et al. (2018). We characterize this behavior using the long-memory parameter, which quantifies the persistence (or anti-persistence) of fluctuations in time series data and is linearly related to the Hurst exponent (H)a widely used measure in the fractal analysis of fMRI signals Wu et al. (2010), Vyushin and Kushner (2009), Dong et al. (2018), Akhrif et al. (2018), Campbell et al. (2022). We extend this modeling to the spatial domain by employing a Bayesian spatial regression framework, enabling voxelwise estimation of long-memory parameters and their relationships with covariates. This spatially informed approach allows us to characterize latent brain dynamics beyond what is captured by connectivity or entropy-based methods.

Long-range dependence, often observed in complex dynamical systems, manifests as scale-free behavior across physical, biological, and social domains Bak et al. (1987). BOLD fMRI signals exhibit similar scale-free properties Ciuciu et al. (2012), where the power spectral density follows a power-law decay, reflecting long-range temporal correlations in the signal. Recent studies Omidvarnia et al. (2022), Medel et al. (2023) have demonstrated consistent associations between brain temporal complexity and the slope of the PSD in log-log space, which is directly related to the long-memory parameter. Consequently, resting-state fMRI signals exhibit long-memory behavior across various brain regions, reflecting intrinsic long-range temporal dependencies in neuronal activity Lahmiri (2018), Campbell and Weber (2022). Investigating these properties is essential for understanding the brains temporal complexity and the mechanisms underlying its functional organization in both health and disease.

We model the autocovariance structure of the fMRI signal at each voxel $v$ using a long-memory process defined by $[\mathbf{\Sigma}_v]_{ij} = \gamma(|i-j|)$, $i$ and $j$ being temporal coordinates, and $\gamma(h) \propto h^{-\alpha}$ for large lag values $h$. The parameter $\alpha$ characterizes the rate of decay and is referred to as the long-memory parameter. This formulation stands in contrast to the commonly assumed autoregressive model of order one (AR(1)), $\gamma(h) \propto \rho^h$, which captures only short-range dependencies and may be inadequate for modeling the persistent correlations observed in resting-state fMRI signals. To analyze these long-range dependencies more effectively, we employ the Discrete Wavelet Transform (DWT), which transforms the fMRI signals into the time-frequency domain. This yields a diagonal covariance matrix $\Sigma^*_v$, whose entries vary with scale and take the form $\Sigma^*_v = v \cdot \mathrm{diag}(2^{-\alpha_v m})$, where $m$ indexes the wavelet scale. This representation provides a more tractable and accurate framework for estimating voxelwise long-memory parameters.

Accurate estimation of the long-memory parameter $\alpha$ is crucial for capturing the underlying temporal dynamics of fMRI signals and relating them to clinical or developmental covariates. To achieve this, we employ Bayesian inference in conjunction with the DWT, which provides both computational efficiency and a probabilistic framework for incorporating prior information and quantifying uncertainty in voxelwise estimates.

There are multiple approaches in the literature for estimating the long-memory parameter. Heuristic and graphical methods, such as the R/S method and Detrended Fluctuation Analysis (DFA), offer straightforward computation of scaling exponents but may be sensitive to noise or limited time series length in practical applications Lo (1991), Peng et al. (1994). In contrast, we adopt a Bayesian approach, which enables coherent uncertainty quantification for the long-memory parameter Jeong et al. (2013). A major computational challenge in voxelwise fMRI data arises from the estimation of high-dimensional and dense covariance matrices. To address this, we employ the Discrete Wavelet Transform (DWT), which decomposes the data into multiple scales and transforms the covariance structure into a near-diagonal form Fadili and Bullmore (2002), thereby substantially reducing computational burden. Following Jeong et al. (2013), we use the variance progression formula Wornell (1993), Vannucci et al. (1998), which models how the variance of wavelet coefficients changes across scales $m$ and directly informs the diagonal structure of $\Sigma_v^*$ used to estimate the long-memory parameter. By estimating the long-memory parameter embedded in the wavelet coefficients, we are able to investigate scaling properties and relate them to underlying neural mechanisms.

Our study makes several key contributions to the field of neuroimaging data analysis. First, we introduce a novel Bayesian spatial modeling approach for estimating long-memory parameters, which addresses the computational challenges posed by high-dimensional covariance structures in voxelwise fMRI data. Second, our analysis provides a comprehensive investigation of long-memory dynamics in fMRI signals, exploring their associations with individual-level covariates. Finally, we identify significant relationships between the long-memory parameter and the age of children and adolescents in the Hippocampus and anterior lobe regions. These findings contribute to the growing body of evidence linking brain temporal complexity to developmental changes in functional brain organization.

The paper is organized as follows. Section 2 outlines the main contribution of this article, our novel methodological approach, including the estimation of individual long-memory maps in Subsection 2.1 and the subsequent multi-subject analysis and dimension reduction in Subsection 2.2, which investigate associations between long-memory parameters and selected covariates. In Section 4 we apply our propose neuroimaging analysis framework to the ADHD 200 data and presents the results of this analysis, summarizing the estimated long-memory maps and identifying subject characteristics associated with the long-memory parameter. Finally, Section 5 discusses the implications of these findings, highlights study limitations, and proposes directions for future research.

## 2 | METHODS

Our proposed pipeline consists of two main stages. In the first stage, we perform a subject-level analysis to estimate long-memory parameters at each voxel. Without loss of generality, we assume that the BOLD time series at each voxel is centered at zero, and model it as a mean-zero Gaussian process with covariance matrix $\boldsymbol{\Sigma}_v$, characterized by a power-law decay. We apply the Discrete Wavelet Transform (DWT) to each voxelwise time series, which transforms the covariance $\boldsymbol{\Sigma}_v$ into a near-diagonal matrix $\boldsymbol{\Sigma}_v^*$, whose structure is governed by two parameters: the long-memory parameter $\alpha_v$ and the innovation variance $\sigma_v^2$. This transformation enables efficient estimation of $\alpha_v$ and $\sigma_v^2$. A Bayesian Markov Chain Monte Carlo (MCMC) method is used to obtain posterior distributions of these parameters at each voxel and for each subject, yielding subject-specific spatial maps of long-memory characteristics.

After estimating the long-memory parameter maps for each subject, we proceed to the second stage of the analysis, which relates these spatial maps to covariates of interest (e.g., clinical or demographic variables). For subject $i$, let $\boldsymbol{\alpha}_i = \{\alpha_{iv}; v = 1, \ldots, N_v\}$ denote the voxelwise long-memory parameter map obtained in Stage 1. We treat $\boldsymbol{\alpha}_i$ as an imaging phenotype summarizing the spatial distribution of temporal dependence across the brain.

To address the challenges of high dimensionality and spatial correlation, we implement a composite-basis approach. This method applies singular value decomposition (SVD) to localized features within a parcellated spatial representation and subsequently captures global spatial patterns. The resulting dimension-reduced spatial features are incorporated into a group-level Bayesian regression model to assess how long-memory properties vary with individual characteristics.

## 2.1 Subject-level analysis

At the subject level, we estimate long-memory parameters independently for each voxel time series. Let $\boldsymbol{Y}_v = (Y_{v1}, \ldots, Y_{vT})^\top$ denote the BOLD signal vector at voxel $v$ with $T$ time points. Assuming the data have been centered, we model $\boldsymbol{Y}_v$ as a realization from a zero-mean Gaussian process with covariance matrix $\boldsymbol{\Sigma}_v$, which reflects the temporal dependence structure at that voxel. To account for this temporal dependence, we model each time series as a long-memory process. These signals exhibit a complex structure characterized by dependencies across multiple time scales.

Figure 1 illustrates the characteristic long-memory behavior of a single fMRI time series from our motivating application, shown in both the temporal and spectral domains. Panel (a) presents a time series of fMRI data acquired from a single voxel, measuring the intensity of the BOLD signal at each time point. Panel (b) depicts the power spectral density (PSD) of this signal for frequencies below 0.5 Hz. The presence of peaks in the PSD indicates dominant frequency components below 0.15 Hz. In panel (c), the log-log plot of spectral density versus frequency shows a flat line over the interval [–3, –0.8], characteristic of white noise behavior. Over the interval [–0.8, –0.1], we observe an approximately linear relationship, which indicates the presence of long-range dependencies in the data. This linear behavior in the log-log plot is a signature of a long-memory process, with the slope often referred to as the spectral exponent directly related to the long-memory parameter Clauset et al. (2009).

A flat log spectral density in the lower log-frequency range implies that the fMRI signal behaves like white noise, whereas the presence of dominant low-frequency components and linearity at higher log frequencies suggests power-law behavior. The Discrete Wavelet Transform (DWT) is particularly suited to address the multi-scale nature of the signal, providing a framework to capture long-memory dynamics efficiently across different frequency bands.

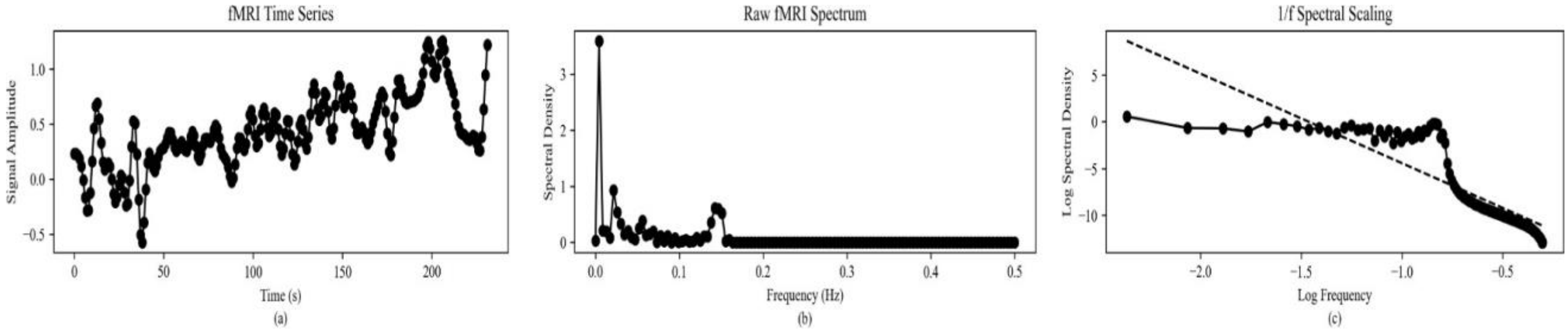

**FIGURE 1** Long-memory behavior of fMRI time series. (a) fMRI time series from a single brain voxel; (b) power spectral density (PSD) of the time series, highlighting dominant low-frequency components; (c) log-log plot of the spectral density, displaying white noise behavior at lower frequencies and long-memory behavior at higher frequencies, characterized by a power-law exponent.

Vannucci et al. (1998) proposed a method to compute the covariance of wavelet coefficients using the recursive filters of the DWT. We adopt their approach to characterize long-memory processes by modeling the covariance structure of the error process $\boldsymbol{E}_v$ at each voxel. Specifically, we assume that the covariance matrix $\boldsymbol{\Sigma}_v(i,j)$ is given by $\gamma(|i-j|)$, where $\gamma(h)$ follows a power-law decay:

$$\gamma(h) \propto \frac{1}{h^{\alpha}}, \tag{1}$$

where $h$ is the lag and $\alpha$ is the long-memory parameter governing the rate of decay of autocorrelations as $h$ increases.

To simplify the dense structure of the autocovariance matrix of the long-memory process, we apply the discrete wavelet transform (DWT) to the processes $\{\boldsymbol{E}_v, v = 1, \ldots, N_v\}$ Mallat (1989). The DWT is based on wavelet basis functions derived from a mother wavelet $\psi$, defined as

$$\psi_{m,n}(x) = 2^{m/2}\psi(2^m x - n),$$

where $m$ denotes the scale (which controls the frequency band analyzed), and $n$ denotes the translation (which determines the location in time). The DWT projects each time series onto this set of basis functions, yielding

$$\boldsymbol{Y}_v^* = \boldsymbol{W}\boldsymbol{Y}_v,$$

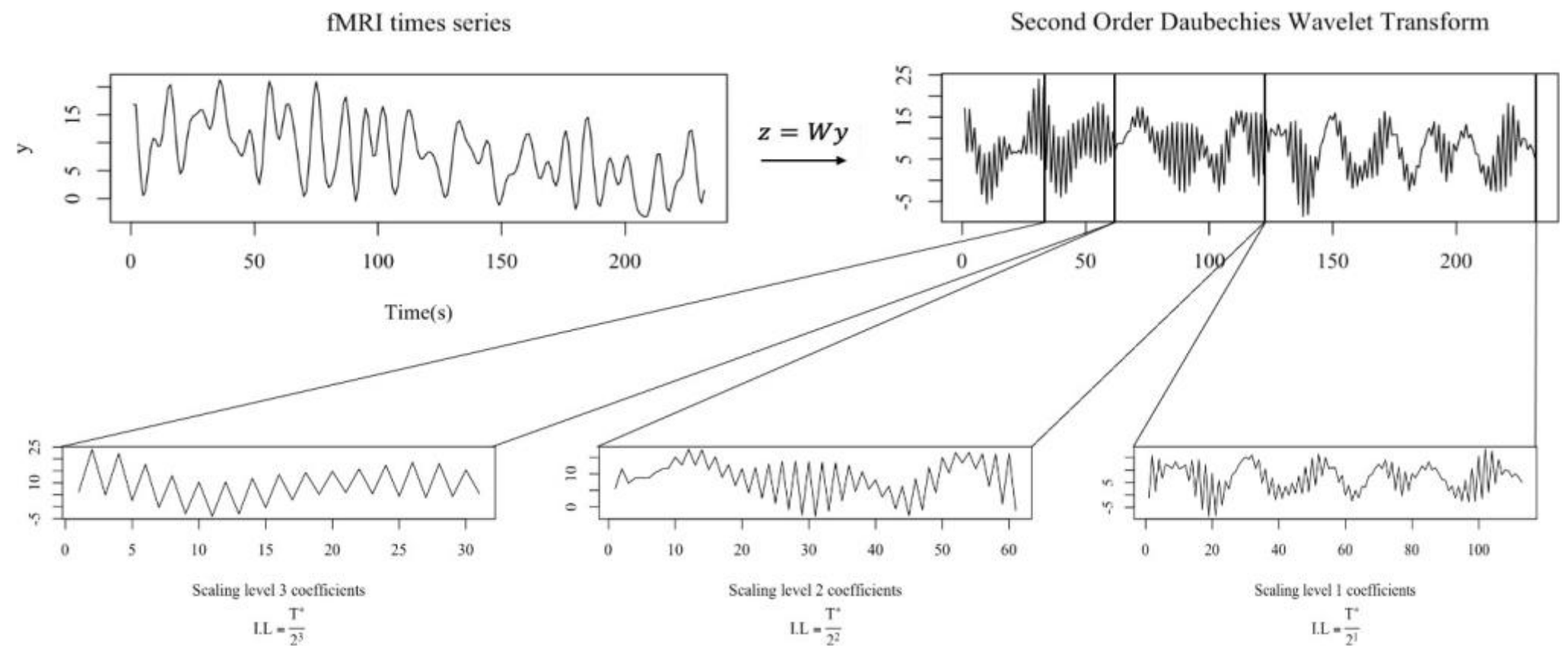

**F I G U R E 2** Top left: time series plot of the fMRI BOLD signal in the time domain. Top right: one-dimensional discrete wavelet transform (DWT) of the signal in the frequency domain. Bottom row: zoomed-in plots of the decomposed signal at different scale levels. The interval length (I.L.) is shown for each scale $j$, defined as $T^*/2^j$, where $T^*$ is the length of the transformed signal.

where the discrete wavelet transformation matrix $\boldsymbol{W}$ is constructed as follows:

1. **Generate wavelet basis functions**: For each scale level $m$ and each translation $n$, the wavelet basis function $\psi_{m,n}(x)$ is computed. This results in a collection of wavelet basis functions that together decompose the signal.
2. **Construct the matrix**: The matrix $\boldsymbol{W}$ is constructed by placing each of these discretized wavelet basis functions as rows. Each row $\boldsymbol{w}_r$ corresponds to a discretized version of the wavelet basis function $\psi_{m,n}(x)$. The rows are ordered by scale, from the finest (capturing high-frequency components) to the coarsest (capturing low-frequency components), and are followed by the scaling function coefficients $\boldsymbol{V}_m$.
3. **Matrix dimensions**: If the original time series contains $T$ time points, then the resulting matrix $\boldsymbol{W}$ has dimensions $T^* \times T$, where $T^*$ depends on the number of wavelet decomposition levels. Each row of $\boldsymbol{W}$ corresponds to a wavelet coefficient associated with a specific scale and time shift, while each column represents the contribution of the original time-domain data to the wavelet coefficients.
4. **Orthogonality**: When orthogonal wavelet basis functions are used, the matrix $\boldsymbol{W}$ is orthogonal, satisfying $\boldsymbol{W}^T\boldsymbol{W} = \boldsymbol{I}$. This property simplifies both the wavelet decomposition (via multiplication by $\boldsymbol{W}$) and the reconstruction of the original signal (via multiplication by $\boldsymbol{W}^T$).

When determining the discrete wavelet transformation matrix, the scaling level $m$ must be specified in advance. A higher value of $m$ increases the ability to capture lower-frequency components of the signal. Therefore, $m$ should be carefully selected based on the frequency range of the data. Figure 2 illustrates a second-order Daubechies wavelet decomposition of a single fMRI time series. The transformed signal is separated into components corresponding to different scale levels, each capturing activity at a distinct frequency range. The figure shows both the overall decomposition and zoomed-in views of selected scale-level components.

When the DWT is applied to $\boldsymbol{Y}_v$, it transforms the covariance matrix from $\boldsymbol{\Sigma}_v$ to $\boldsymbol{\Sigma}_v^* = \boldsymbol{W}\boldsymbol{\Sigma}_v\boldsymbol{W}^T$. This transformation yields the covariance matrix of the time series $\boldsymbol{Y}_v^*$ in the wavelet domain for each voxel $v$. The correlation between wavelet coefficients at different scales decreases rapidly Dijkerman and Mazumdar (1994), Fan (2003). Therefore, it is reasonable to approximate $\boldsymbol{\Sigma}_v^*$ as a diagonal covariance matrix, with diagonal elements $\nu_v\sigma^2_{mn}$ corresponding to the variance of the $n$th wavelet coefficient at the $m$th scale level. We adopt the variance progression formula proposed in Wornell (1993), Jeong et al. (2013):

$$\nu_v\sigma^2_{mn} = \nu_v(2^{\alpha_v})^{-m}, \tag{2}$$

where $\nu_v$ is the innovation variance, $\alpha_v \in (0, 1)$ is the long-memory parameter, and $m$ is the wavelet scale level. Here larger values of m indicate coarser approximations. The long-memory parameter $\alpha_v$ is linearly related to the Hurst exponent $H$ by the equation $\alpha_v = 2 - 2H$ for $1/2 < H < 1$, as established in Vyushin and Kushner (2009), Suibkitwanchai (2022).

We estimate the long-memory parameter $\alpha_v \in (0, 1)$ simultaneously with the innovation variance $v_v$ using a Bayesian approach, assuming $\alpha_v \sim \text{Beta}(a, b)$ and $v_v \sim \text{Inv-Gamma}(p, s)$ *a priori*. The Markov chain Monte Carlo (MCMC) algorithm used to sample from the posterior distribution is described in Appendix A.1. Our approach combines a truncated MetropolisHastings algorithm to sample from the posterior of $\alpha_v$ with Gibbs sampling for $v_v$. This yields posterior distributions of the long-memory parameters across all voxels for each subject.

## 2.2 | Group-level analysis

For subject $i$, let $\alpha_{i,v}$ denote the long-memory parameter at voxel $v$, for $i = 1, \ldots, N$. Let $\boldsymbol{\alpha}_v = (\alpha_{1,v}, \ldots, \alpha_{N,v})^\top$ denote the vector of estimated posterior means of the long-memory parameters at voxel $v$ across all $N$ subjects. We model $\boldsymbol{\alpha}_v$ as a function of subject-specific covariates as follows:

$$\boldsymbol{\alpha}_v = \boldsymbol{Z}\boldsymbol{\beta}_v + \boldsymbol{\epsilon}_v, \tag{3}$$

where $\boldsymbol{Z}$ is an $N \times Q$ design matrix. The regression coefficients $\boldsymbol{\beta}_v$ characterize the relationship between the covariates $\boldsymbol{Z}$ and the estimated long-memory parameters at voxel $v$. Collectively, we define $\boldsymbol{\beta} = \{\boldsymbol{\beta}_1, \boldsymbol{\beta}_2, \ldots, \boldsymbol{\beta}_{N_v}\}$, representing the regression coefficients across all voxels, for $v = 1, 2, \ldots, N_v$.

We apply the composite-basis approach proposed in Miranda and Morris (2021) to capture both local and global spatial dependencies across voxels in the regression matrix $\boldsymbol{\beta}$, while effectively reducing dimensionality to maintain computational tractability. The voxelwise long-memory parameters $\alpha_{i,v}$ obtained from the subject-level analysis characterize temporal dependence within individual BOLD time series. However, these parameters form spatial maps across the brain, and spatial dependencies among voxels are explicitly modeled at the group level through the composite-basis framework. In this way, the composite-basis structure provides the mechanism by which spatial dependencies are captured in the regression model.

The composite-basis approach consists of two steps. First, the brain is divided into biological regions of interest (ROIs). For each ROI, local bases are obtained using singular value decomposition (SVD) of data matrices composed of observations within that ROI, capturing structured variation among nearby voxels. The data are then projected onto these local bases. Second, the projected data are used to derive a second-level basis via SVD applied to the output of the first step, thereby capturing spatial relationships across ROIs at a global scale.

Let $\boldsymbol{\alpha}$ denote an $N \times N_v$ matrix, where each row $i$ contains the long-memory map for subject $i$ across all voxels, i.e., the row vector $(\alpha_{i,1}, \ldots, \alpha_{i,N_v})$. We apply the composite-basis approach by rewriting the model in (3) as follows:

$$\boldsymbol{\alpha}\mathbf{Y} = \boldsymbol{Z}\boldsymbol{\beta}\mathbf{Y} + \boldsymbol{\epsilon}\mathbf{Y}. \tag{4}$$

where $\mathbf{Y} = \boldsymbol{\Phi}\boldsymbol{\Psi}$ denotes the composite basis. The matrix $\boldsymbol{\Phi} = \text{blkdiag}(\Phi_1, \ldots, \Phi_{N_{\text{ROI}}})$ is block-diagonal, with each block $\Phi_r$ consisting of the orthonormal eigenvectors obtained from the singular value decomposition of the data in the $r$th ROI. These matrices capture the dominant modes of variability within each ROI. The second-level basis $\boldsymbol{\Psi}$ captures variability across ROIs, based on the projected local features. This approach first projects each ROI into a lower-dimensional space containing local features, and then captures global structure using the basis $\boldsymbol{\Psi}$, whose columns are orthonormal eigenvectors of the projected data. The resulting composite basis provides regularization and dimensionality reduction by thresholding the number of principal components according to the proportion of explained variability.

Model (4) can be rewritten as

$$\boldsymbol{\alpha}^* = \boldsymbol{Z}\boldsymbol{\beta}^* + \boldsymbol{\epsilon}^*, \tag{5}$$

where $\boldsymbol{\alpha}^*$ is a $N \times PC_f$ matrix containing the long-memory maps projected into the basis space. The matrix $\boldsymbol{Z}$ is the $N \times Q$ design matrix containing population-level covariates. For each column of $\boldsymbol{\epsilon}^*$, we assume $\epsilon^*_r \sim N(0, \boldsymbol{\delta}^2_r \boldsymbol{I})$, for $r = 1, \ldots, PC_f$. The $Q \times PC_f$ coefficient matrix $\boldsymbol{\beta}^*$ captures the relationship between the long-memory maps and the covariates of interest.

To estimate $\boldsymbol{\beta}^*$, we specify the following priors:

$$\begin{aligned} \boldsymbol{\beta}^*_r \mid \boldsymbol{\delta}^2_r &\sim \text{MVN}(\mu_0, \delta^2_r \Lambda_0), \\ \boldsymbol{\delta}^2_r &\sim \text{Inverse-Gamma}(k, l) \end{aligned} \tag{6}$$

Note that the prior for $\boldsymbol{\beta}^*_{PC_r}$ depends on $\boldsymbol{\delta}^2_{PC_r}$. The hyperparameters for the inverse-gamma distribution were set to $(k, l) = (0.1, 0.5)$ to restrict large variances in the regression coefficients. For the multivariate normal distribution, we defined the hyperparameters as $(\mu_0, \Lambda_0) = (\mathbf{0}, 100 \times (\boldsymbol{Z}^\top \boldsymbol{Z})^{-1})$, based on the assumption that the estimated values of $\boldsymbol{\beta}^*$ would be centered around zero. Estimation of $\boldsymbol{\beta}^*$ proceeds via MCMC and is described in Appendix A.2. After obtaining posterior samples for $\boldsymbol{\beta}^*$, we project these samples back into the original voxel space via the inverse of the composite basis, $\mathbf{Y}^{-1}$. These steps enable voxel-level inference even though the MCMC sampling is done over the lower-dimensional basis space.

## 3 | SIMULATION STUDY

We conducted a comprehensive multi-subject simulation study to evaluate the performance and robustness of the proposed wavelet-based Bayesian framework for estimating voxelwise long-memory (LM) parameters. Subject-specific spatial maps of the LM parameter were simulated and wavelet coefficients were generated according to the variance progression model underlying our likelihood, ensuring consistency between signal generation and estimation.

To mimic the biological complexity of rs-fMRI data, we construct latent $\alpha$-maps on a 2D gray-matter mask and introduce multi-subject heterogeneity. Each subject's map incorporates global and region-specific perturbations of the $\alpha$ parameter, alongside voxel-level noise, to reflect realistic between-subject variability. We also impose spatial correlation within these maps using a relaxed Gaussian Markov Random Field (GMRF) smoothing approach, which introduces local dependence while preserving the boundaries of simulated regions of interest. All simulated datasets are integrated with the AAL atlas and associated lookup tables, making them fully compatible with the two-stage analysis and dimensionality reduction pipeline described in Section 2.2 and Equation 3.

**Scenario 1.** As an initial validation experiment, we consider a spatial multi-region simulation designed to assess recovery of long-memory parameters across a broad dynamic range. This setting provides a controlled baseline for evaluating estimation accuracy when LM values span nearly the full admissible parameter space, and serves to characterize overall performance of the proposed framework before introducing additional complexity. Figure 3 shows the true and estimated LM maps averaged across 100 simulated subjects, illustrating close agreement within activated regions. The visual correspondence across low and high LM regions demonstrates that the proposed estimator remains stable across the entire parameter range, without systematic distortion at boundary values. Region-specific estimation accuracy is summarized in Table 1 using MSE, RMSE, and bias averaged across subjects.The consistently small bias and low RMSE across all regions align with what we see visually in Figure 3.

| Circle Position | $n_{\text{voxels}}$ | MSE | RMSE | Bias |
|---|---|---|---|---|
| LL | 113 | 0.00169 | 0.04108 | 0.04061 |
| LR | 117 | 0.00059 | 0.02425 | 0.01935 |
| UL | 119 | 0.00027 | 0.01638 | 0.00970 |
| UR | 120 | 0.00020 | 0.01405 | 0.00659 |

**TABLE 1** Results for simulation Scenario 1. Region-specific error estimation for the baseline multi-region simulation. Mean squared error (MSE), root mean squared error (RMSE), and bias are reported for each activated region, averaged across simulated subjects.

**Scenario 2.** To examine robustness to innovation variance, we vary $v$ over a grid spanning [0.5, 25] while keeping all other aspects of the simulation fixed. This scenario evaluates whether changes in noise scale impact recovery of the LM parameter. Figure 4 reports mean accuracy metrics across groups of 25 subjects for each subscenario, indicating that the model remains robust to innovation variance modifications. Specifically, the stability of MSE and bias across the $v$ grid confirms that scaling the variance does not confound estimation of the LM parameter.

**Scenario 3.** We further examine sensitivity to detect intensity differences by varying the LM level within a single activated region while keeping the spatial configuration fixed. Specifically, the lower-left circle was assigned values spanning [0.05, 0.95] by 0.1, enabling us to investigate the minimum detectable intensity difference, particularly at the second-stage analysis. As shown in Figure 5, our two-stage Bayesian model reliably detects differences as small as 0.1 in the lower-left region, with estimated group-level LM contrasts closely matching the true simulated effect, while frequentist results are provided for comparison. Both images has been corrected for FDR at significance level of 0.05. These findings demonstrate that the proposed framework preserves sensitivity to subtle LM contrasts while maintaining spatial specificity under multiplicity correction.

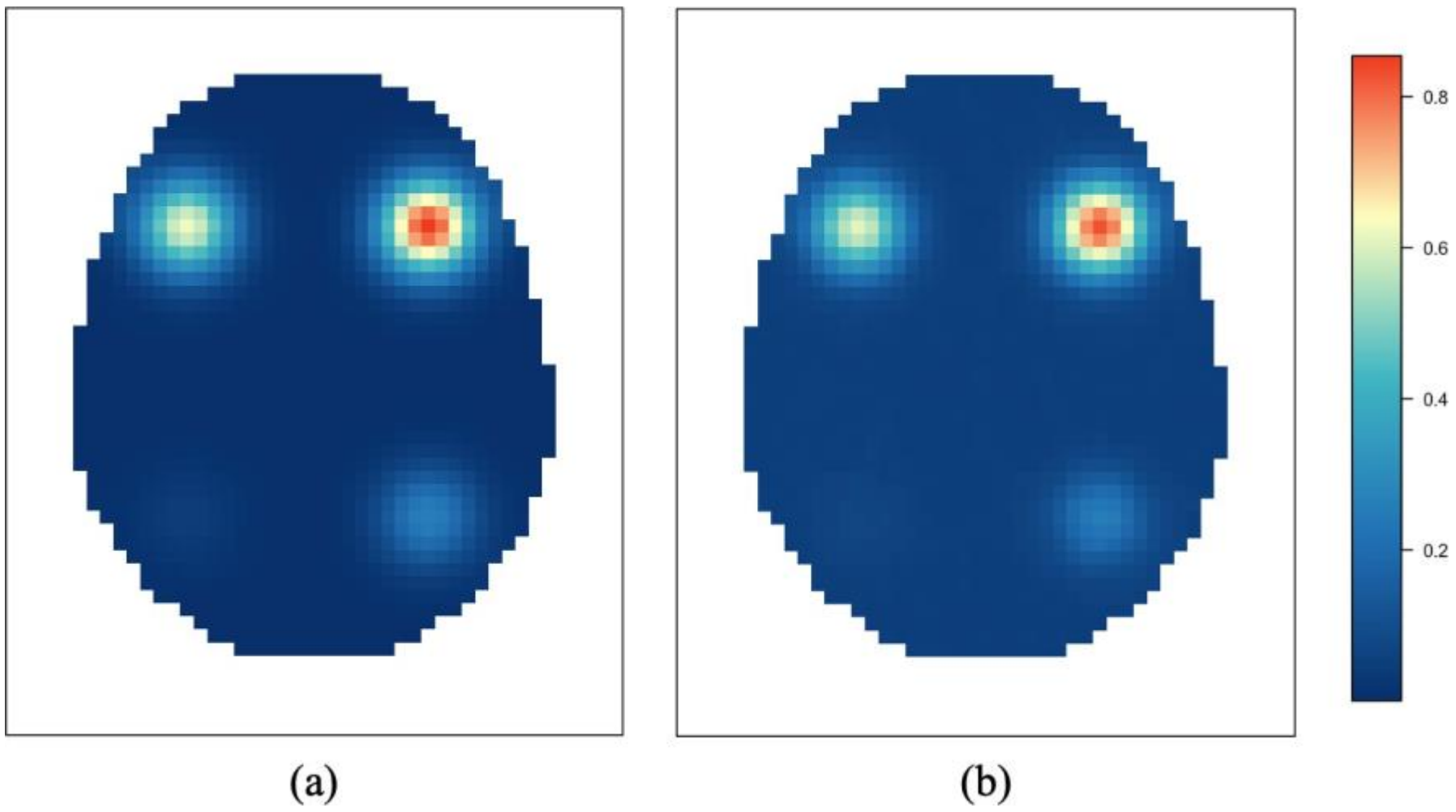

**FIGURE 3** Simulation Scenario 1. Voxelwise averages of simulated long-memory parameter maps (a) True LM map; (b) mean estimated LM map.

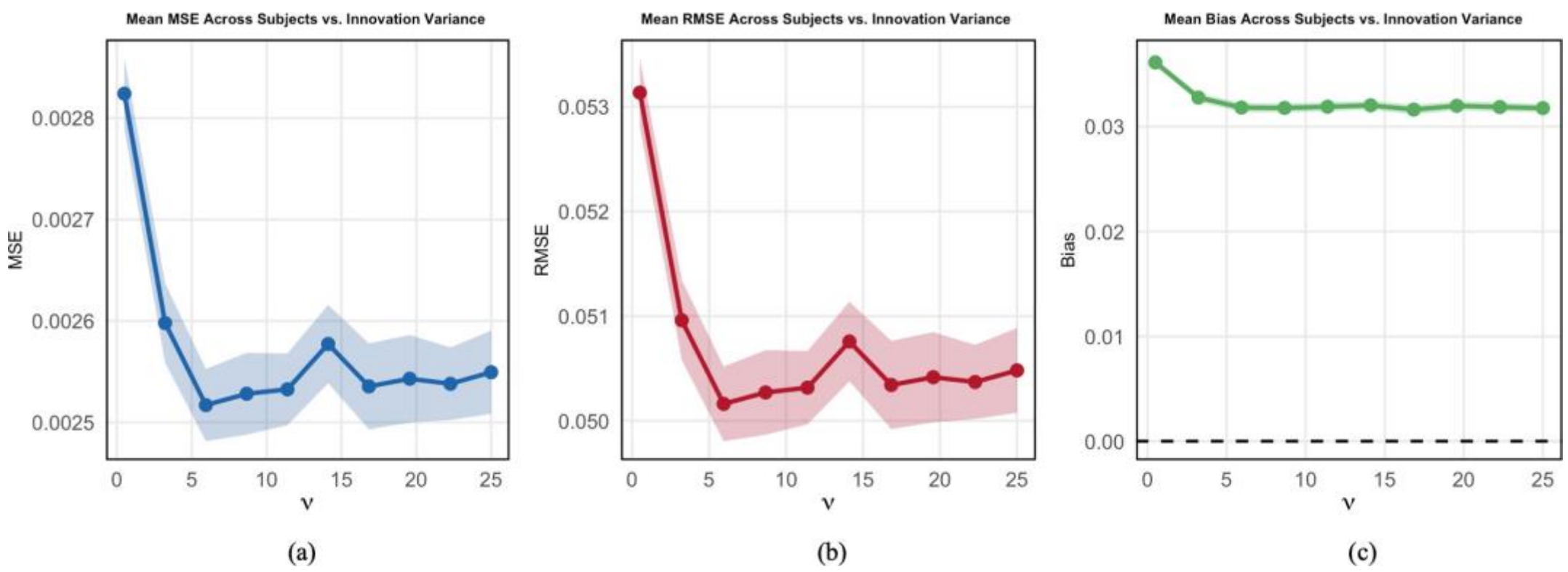

**FIGURE 4** Simulation Scenario 2. Estimation accuracy of LM parameter under varying innovation variance; mean (a) MSE; (b) RMSE; (c) bias, with shaded bands indicating 95% confidence intervals.

**Scenario 4.** To assess sensitivity to spatial variability, we consider a scenario in which only the size of the activated regions is modified while all other simulation components remain fixed. This setting isolates the effect of region size and provides insight into the ability of the two-stage framework to capture increasingly localized spatial patterns. Figure 6 summarizes estimation accuracy across radius levels, showing increasing MSE and RMSE as regions become smaller, highlighting reduced recovery performance for highly localized activations. Figure 7 further compares group-level effect maps from the two-stage Bayesian and frequentist approaches, illustrating improved localization of small-radius activations under the Bayesian framework. As the activation region becomes very small, it is expected that any statistical model will lose some estimation power due to reduced signal support; however, even at the smallest radius level, the proposed framework was still able to detect the activation, indicating strong sensitivity to localized effects.

**Scenario 5.** To further evaluate robustness under measurement noise, we consider a noise perturbation scenario in which additive Gaussian noise is injected directly into the wavelet coefficients according to

$$z_v \rightarrow z_v + Ce, \quad e \sim \mathsf{N}(0, 1) \tag{7}$$

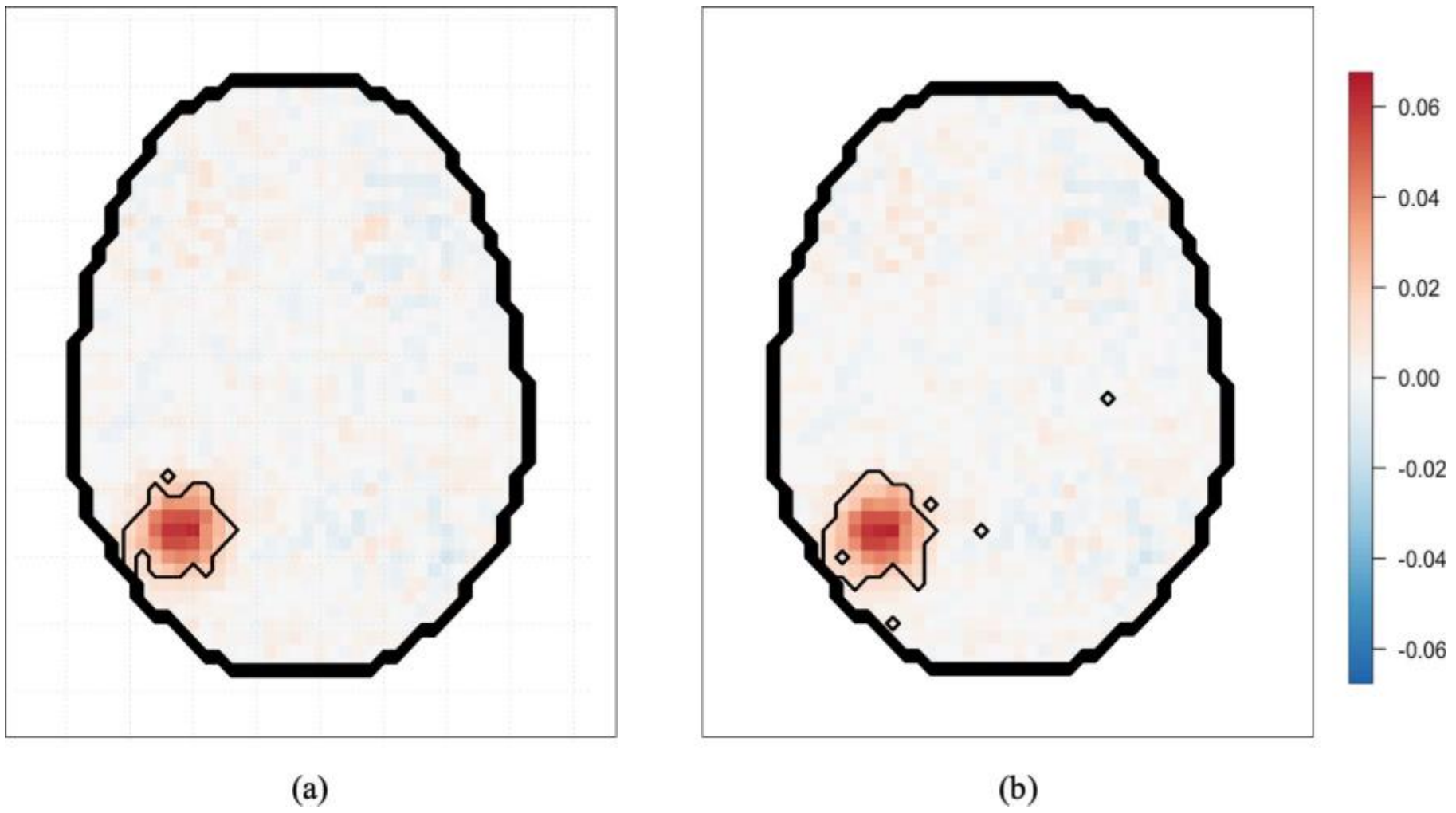

**FIGURE 5** Simulation Scenario 3. Intensity maps of estimated $\beta$ using (a) Two-stage Bayesian method; (b) Frequentist method each corrected for multiple comparisions using an FDR threshold of 0.05.

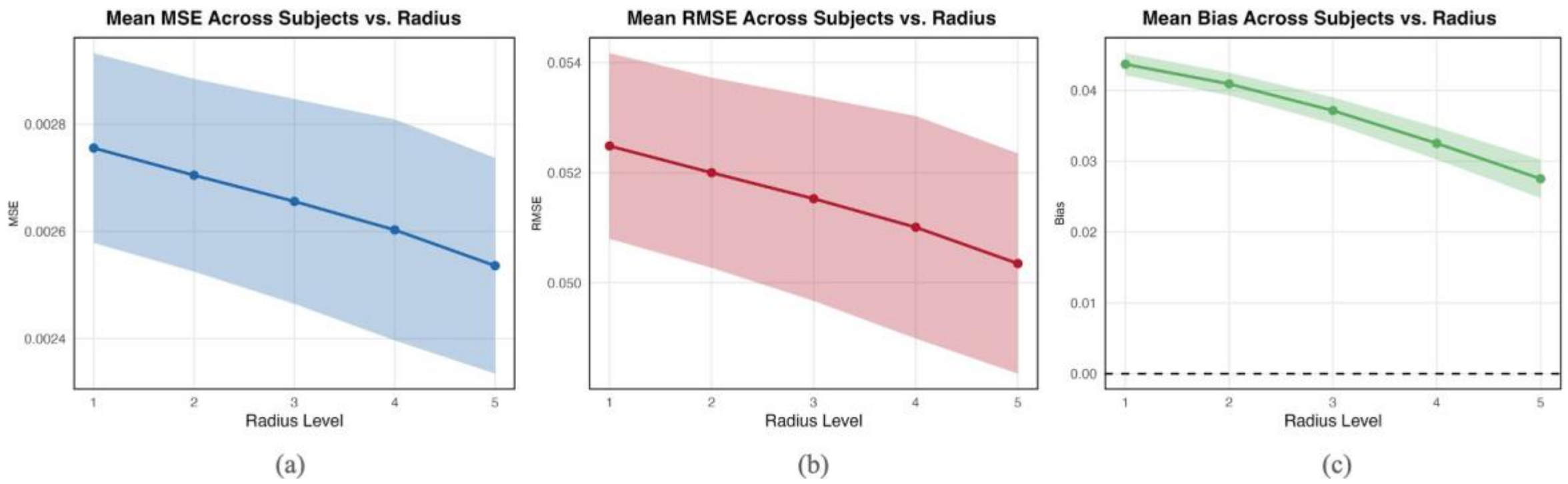

**FIGURE 6** Simulation Scenario 4. Mean estimation accuracy as a function of region radius of LM parameter. Panels show MSE (left), RMSE (middle), and bias (right) averaged across subjects for decreasing activation sizes (radius levels 1–5). Shaded bands indicate 95% confidence intervals.

with scaling factors $C = 0, 0.2, 0.4, 0.6, 0.8, 1$. This setting progressively reduces the signal-to-noise ratio while keeping the underlying spatial LM structure fixed. The goal of this experiment is to assess the stability of LM recovery as noise dominates the observed signal. Figure 8 summarizes estimation accuracy across noise levels, showing increasing MSE and RMSE as $C$ grows, indicating progressive degradation of LM recovery as measurement noise overwhelms the underlying spatial structure. Figure 9 further illustrates group-level $\beta$ maps under most noisy scenario for both four- and two-region settings, comparing the proposed two-stage Bayesian approach with a frequentist alternative. Although increasing noise makes estimation more difficult, the main spatial patterns remain recoverable across moderate noise levels, indicating robustness of the LM estimation procedure.

Finally, we compared the proposed Bayesian hierarchical spatial framework with a standard frequentist voxelwise regression approach on multiple proposed scenarios. Both methods produced highly consistent effect estimates, indicating strong agreement in recovered group-level patterns. Importantly, regions detected by the Bayesian approach were also supported by the frequentist results, suggesting robust overlap between methods.

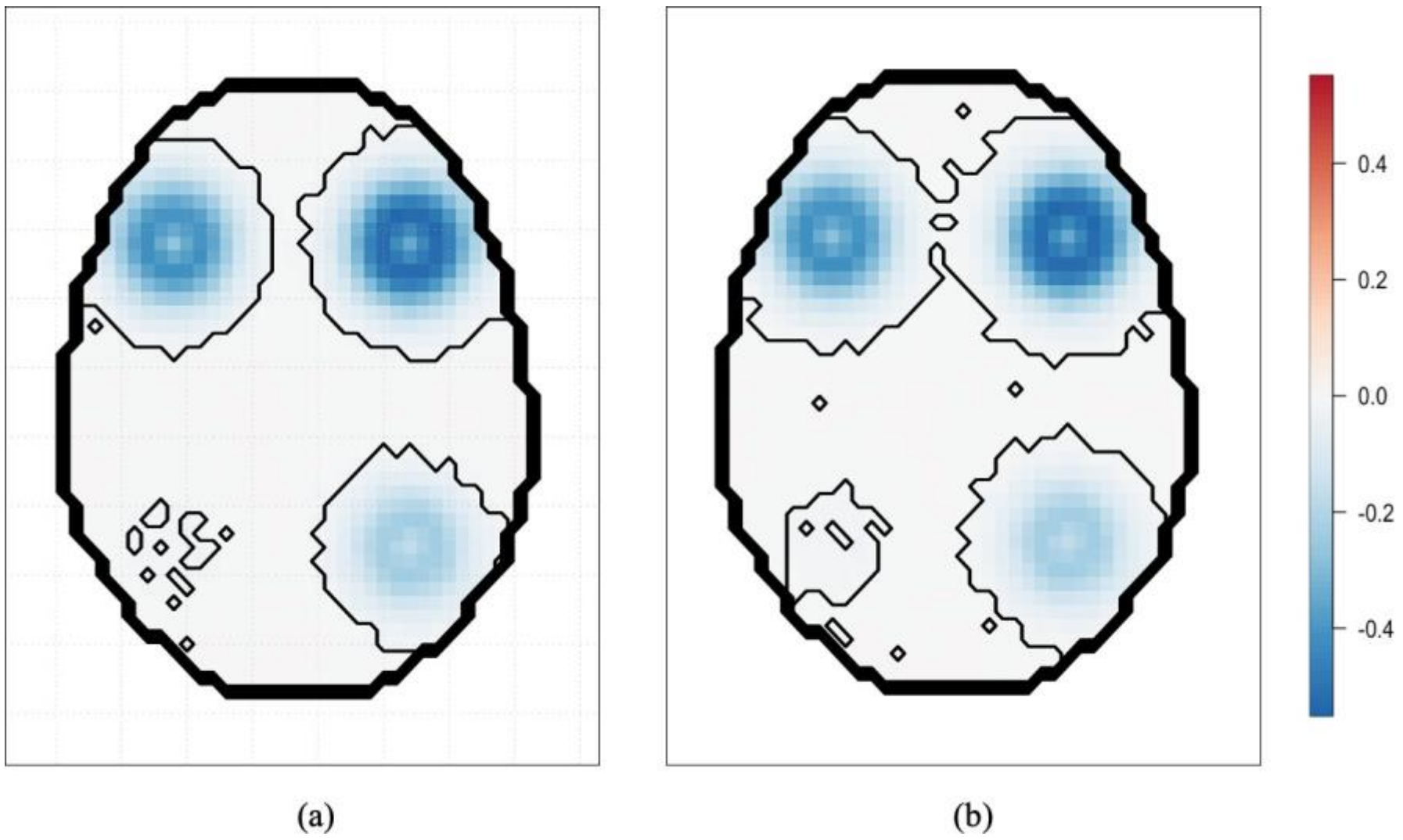

**FIGURE 7** Simulation Scenario 4. Estimated group-level $\beta$ maps under the radius-modification scenario using (a) Two-stage Bayesian method; (b) Frequentist method each corrected for multiple comparisons using an FDR threshold of 0.05.

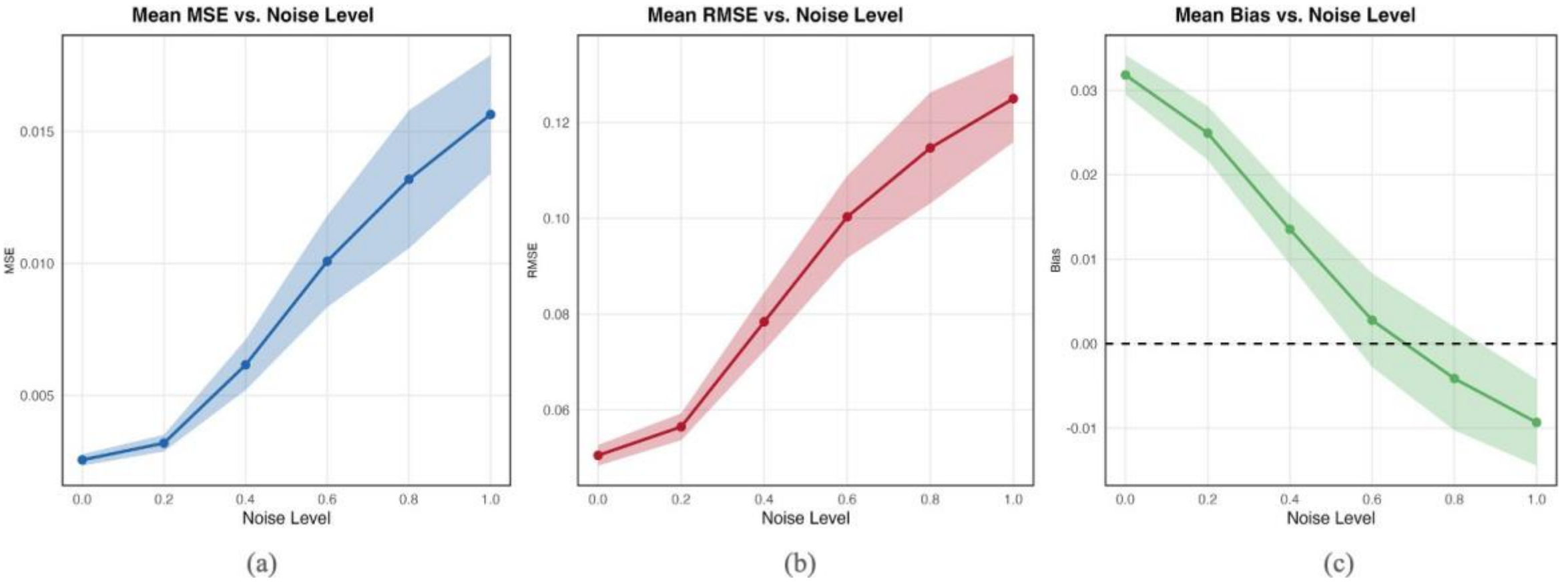

**FIGURE 8** Simulation Scenario 5. Estimation accuracy of LM parameter under increasing noise levels; mean (a) MSE; (b) RMSE; (c) bias, with shaded bands indicating 95% confidence intervals.

## 4 | RESULTS

We analyze data from the ADHD-200 consortium, publicly available through the Neuroimaging Informatics Tools and Resources Clearinghouse (NITRC) and detailed in Bellec et al. (2017). This dataset consists of 973 individuals aged 7 to 27 years with preprocessed resting-state functional MRI (rs-fMRI) data, collected from 17 studies across 8 sites. Of these individuals, 352 are female and the remaining are male. The data include phenotypic information alongside rs-fMRI for each participant.

The repository offers data processed through multiple pipelines: Athena, Burner, and NIAK. We selected the Athena pipeline for this study due to its compatibility with our research objectives. The Athena pipeline processed rs-fMRI using a custom BASH script that integrates AFNI Cox (1996) and FSL Smith et al. (2004) neuroimaging toolkits. Each rs-fMRI dataset consists of 49 Œ 58 Œ 47 voxels and varies in length from 75 to 234 time points due to different acquisition protocols. Preprocessing steps included removal of the first four volumes, slice-timing correction, and motion correction using realignment. The functional data were then co-registered to the standard MNI152 template space with a resolution of 4 Œ 4 Œ 4 mms¸. To mitigate

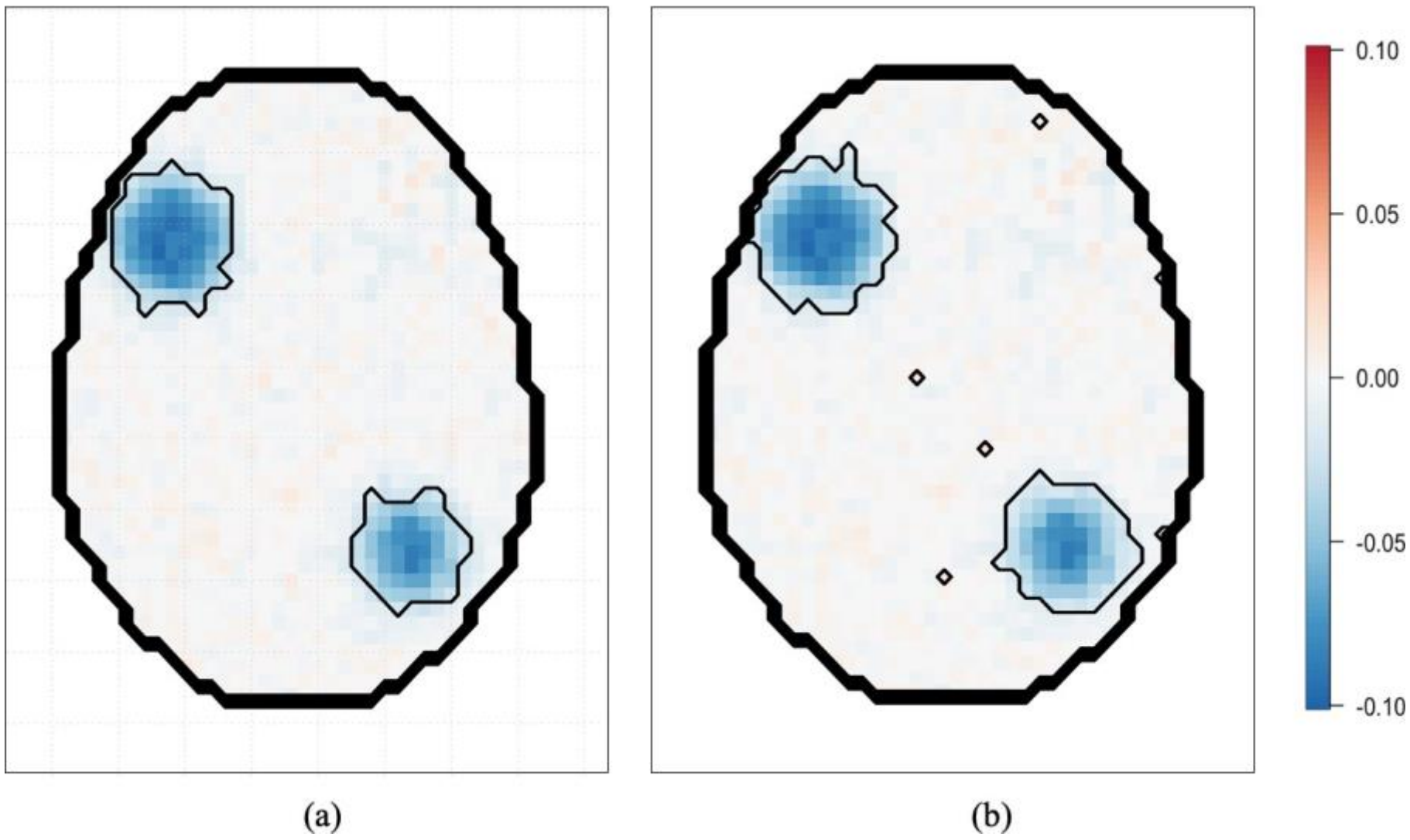

**F I G U R E 9** Simulation Scenario 5. Estimated group-level $\beta$ maps under additive noise scenario using (a) Two-stage Bayesian method; (b) Frequentist method each corrected for multiple comparisons using an FDR threshold of 0.05.

physiological noise, head motion artifacts, and scanner drift, nuisance regressors, including signals from white matter, cerebrospinal fluid, and six head motion parameters, were regressed out. Finally, the data were spatially smoothed with a 6 mm FWHM Gaussian kernel.

We analyzed a final sample of 355 participants with complete unfiltered rs-fMRI data and full phenotypic information for key covariates. This subset was selected to ensure the data were suitable for time-frequency analysis at the first level. Phenotypic variables included sex at birth, age, ADHD diagnosis (combined, inattentive, or hyperactive-impulsive), and medication status. ADHD severity was measured using the Conners Parent Rating ScaleRevised, long version (CPRS-LV) Conners (1997), treated as a continuous variable ranging from 19 to 90. Age was also treated as continuous (range: 718), while sex at birth and medication status were binary. Previous studies have identified associations between ADHD symptoms and age, sex, and medication status Ivanov et al. (2014), Yang et al. (2011), Bollmann et al. (2017). Based on these findings, we selected age, ADHD severity, medication status, and their interaction as primary covariates of interest. Table 2 summarizes the distribution of participants across key subgroups.

To account for differences in temporal sampling across acquisition sites, where repetition times (TRs) were either 2000 or 2500 ms, depending on the acquisition site, rs-fMRI time series with 2500 ms were temporally resampled prior to wavelet decomposition. Specifically, each subject's time series was linearly interpolated to a common TR of 2000 ms. This preprocessing step ensured that each discrete wavelet transform (DWT) scale corresponded to the same frequency band across subjects, thereby enabling meaningful comparisons of wavelet-based long-memory estimates. By aligning the temporal sampling rate across all subjects, this procedure minimizes frequency-scale inconsistencies that can otherwise arise in multi-site datasets with heterogeneous acquisition protocols.

| | ADHD/MED | ADHD/NO MED | NO ADHD |
|---|---|---|---|
| Male | 7 | 24 | 101 |
| Female | 36 | 72 | 115 |

**T A B L E 2** Distribution of participants by sex and ADHD diagnosis. MED: medicated, NO MED: Unmedicated

Our analysis pipeline is summarized in Figure 10. Panel A shows that for each subject $i$, we first estimate long-memory parameter maps $\alpha_v, v = 1, \ldots, N_v$ using the Bayesian approach described in Section 2.1. Panel B shows how these subject-level maps are combined to form a group-level representation of long-memory patterns across the brain. Panel C illustrates how Bayesian inference is then used to examine the effects of covariates on the long-memory parameters in the reduced space as illustrated in part C of Figure 10.

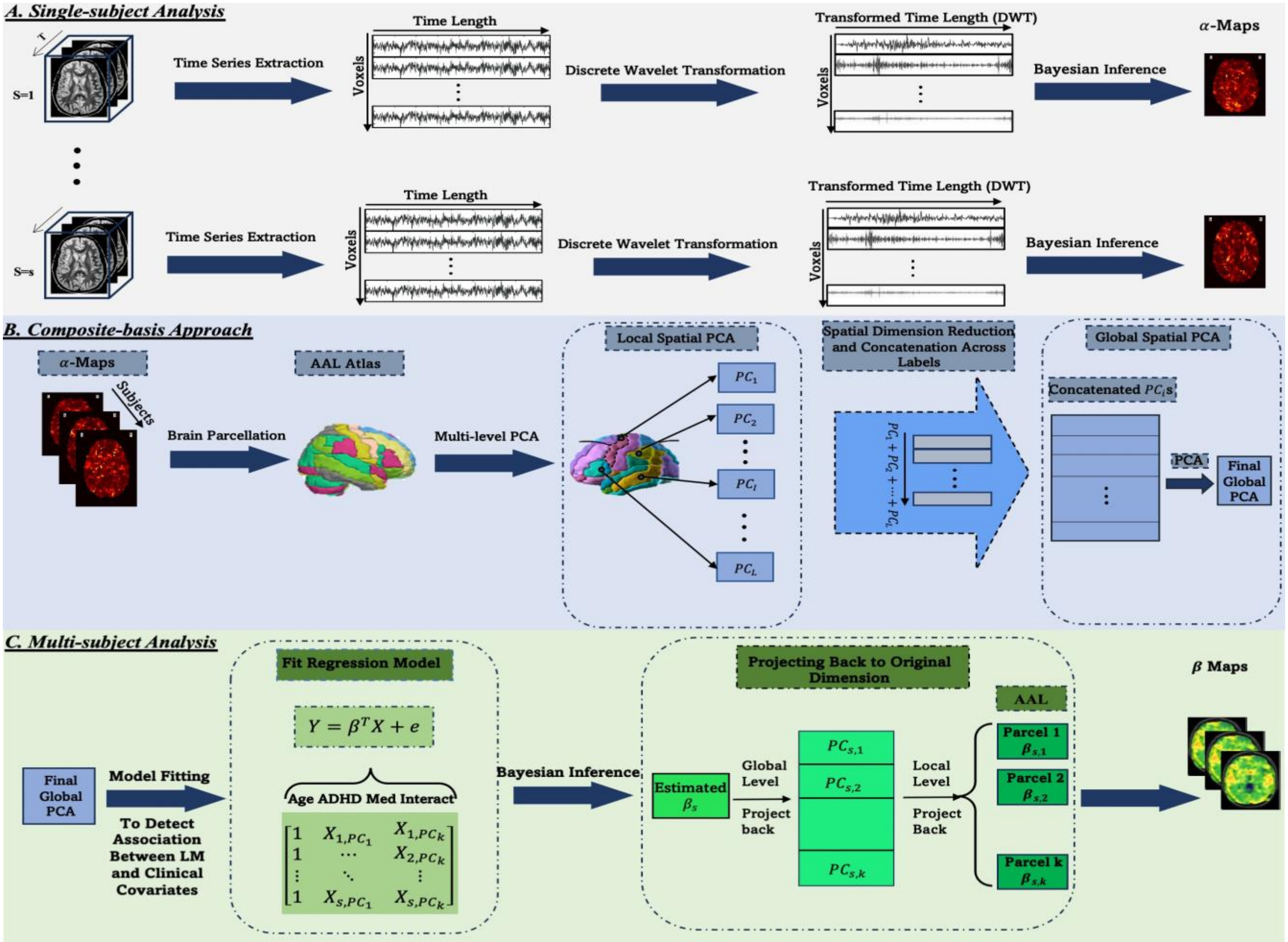

**FIGURE 10** Flowchart of the analysis pipeline. Step A shows the subject-level analysis. First, the Discrete Wavelet Transform is applied to the extracted time series from the 4D fMRI data. Then, long-memory parameters (**α**-maps) are estimated for each subject using the Bayesian approach. Step B shows the group-level analysis. Individual **α**-maps are combined across subjects, and a composite-basis approachconsisting of two successive PCA steps on the spatial dimensionis applied. Finally, Bayesian inference is performed on the reduced data, and the estimated covariate effects are projected back to voxel space.

## 4.1 | Subject-level analysis

At the subject level, we estimated the voxelwise covariance matrix $\mathbf{\Sigma}_v^*$ in the wavelet domain, parameterized by the long-memory parameter $\alpha_v \in (0, 1)$ and the innovation variance $v_v$, as described in Section 2.1. We assigned a Beta(3, 3) prior to $\alpha_v$ and an inverse-gamma(2, 2) prior to $v_v$, with the latter chosen to accommodate large variance values through heavier tails. Estimation was carried out via MCMC sampling, with implementation details provided in Appendix A.

For visualization, we computed voxelwise averages of the estimated subject-level long-memory parameter maps within selected subgroups. Figure 11 presents the voxelwise group-averaged long-memory maps for two representative subgroups selected for illustration purposes: participants aged 12 years or older, and those aged 9 years or younger. In the next section, we examine how covariates influence the spatial distribution of long-memory estimates.

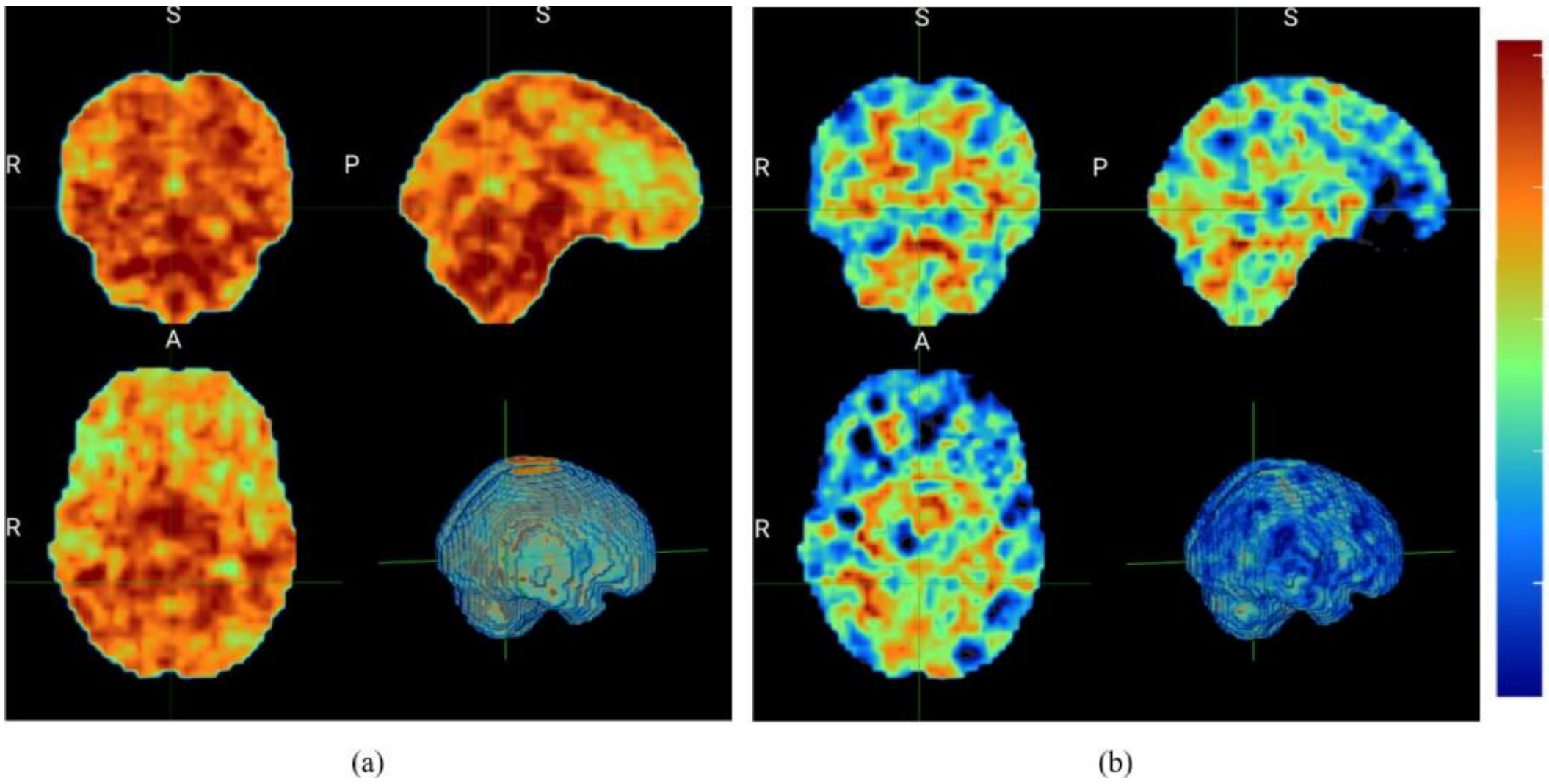

**F I G U R E 11** Voxelwise averages of long-memory maps of (a) 12-year-old or older subjects with no medication; (b) 9-year-old or younger subjects with medication.

## 4.2 | Group-level analysis

We conducted a group-level analysis to examine how long-memory properties of the BOLD signal vary with age, ADHD severity, and medication status. Specifically, we assessed how these covariates relate to spatial patterns in the estimated long-memory parameters across the brain.

We first parcellated the brain into regions of interest (ROIs) using the Automated Anatomical Labeling (AAL) atlas and applied the two-level dimensionality reduction strategy inspired by multi-scale factor analysis (MSFA) Ting et al. (2018). Within each ROI, we used singular value decomposition (SVD) to extract principal components explaining at least 99% of the variance in the projected long-memory values, capturing local spatial structure. We then applied a second-level SVD across the retained components to identify dominant modes of variation across the brain. The projection of the long-memory maps onto the composite basis space reduced the number of spatial locations from 29,639 voxels to 411.

To evaluate whether the observed association between age and the long-memory parameter could be explained by structural brain differences, we included a structural covariate in the Stage 2 design matrix. Morphological differences associated with ADHD may influence spatial normalization and, consequently, affect estimates of temporal autocorrelation through interpolation during image registration. To mitigate this potential source of bias, we extracted the mean grey matter density within the hippocampal region of interest from the grey matter density maps generated by the Athena preprocessing pipeline Bellec et al. (2017) and included it as a covariate in the design matrix. The final design matrix comprised the following predictors: Age (continuous), Medication Status (binary indicator of medication use among participants with ADHD), ADHD Index (a continuous measure of symptom severity derived from the Conners' Parent Rating Scale–Long Version [CPRS-LV]), the ADHD Index $\times$ Medication Status interaction, and Hippocampal Grey Matter Density (continuous), together with an intercept, resulting in a design matrix of dimension $355 \times 6$.

We estimated the regression coefficients using Bayesian inference with priors defined in Equation (6). The inverse-gamma hyperparameters were set to $(k, l) = (0.1, 0.5)$ to discourage high variances, as most variability was already captured through PCA. For the multivariate normal prior, we used $(\mu_0, \Lambda_0) = (0, 0.5 \times I)$. The full conditional distributions of $\beta^*{}_{PC_i}$ and $\delta^2_{PC_i}$ are available in closed form, allowing us to implement posterior sampling using a Gibbs sampler.

The estimated regression coefficients $\boldsymbol{\beta}^*$ form a matrix of dimensions $6 \times 411$, representing the effects of the six covariates on the 411 principal components. To recover voxelwise coefficient maps, we apply a two-step back-projection. First, the coefficients are projected from the global principal component space to the local ROI level. Then, they are multiplied by the corresponding eigenvectors within each ROI to reconstruct voxelwise estimates. This back-projection is performed at each MCMC iteration, yielding a sequence of voxelwise coefficient maps with dimensions $29,639 \times 6$. Because the reconstruction is based on a low-dimensional representation, the resulting maps are spatially regularized and exhibit reduced high-frequency noise.

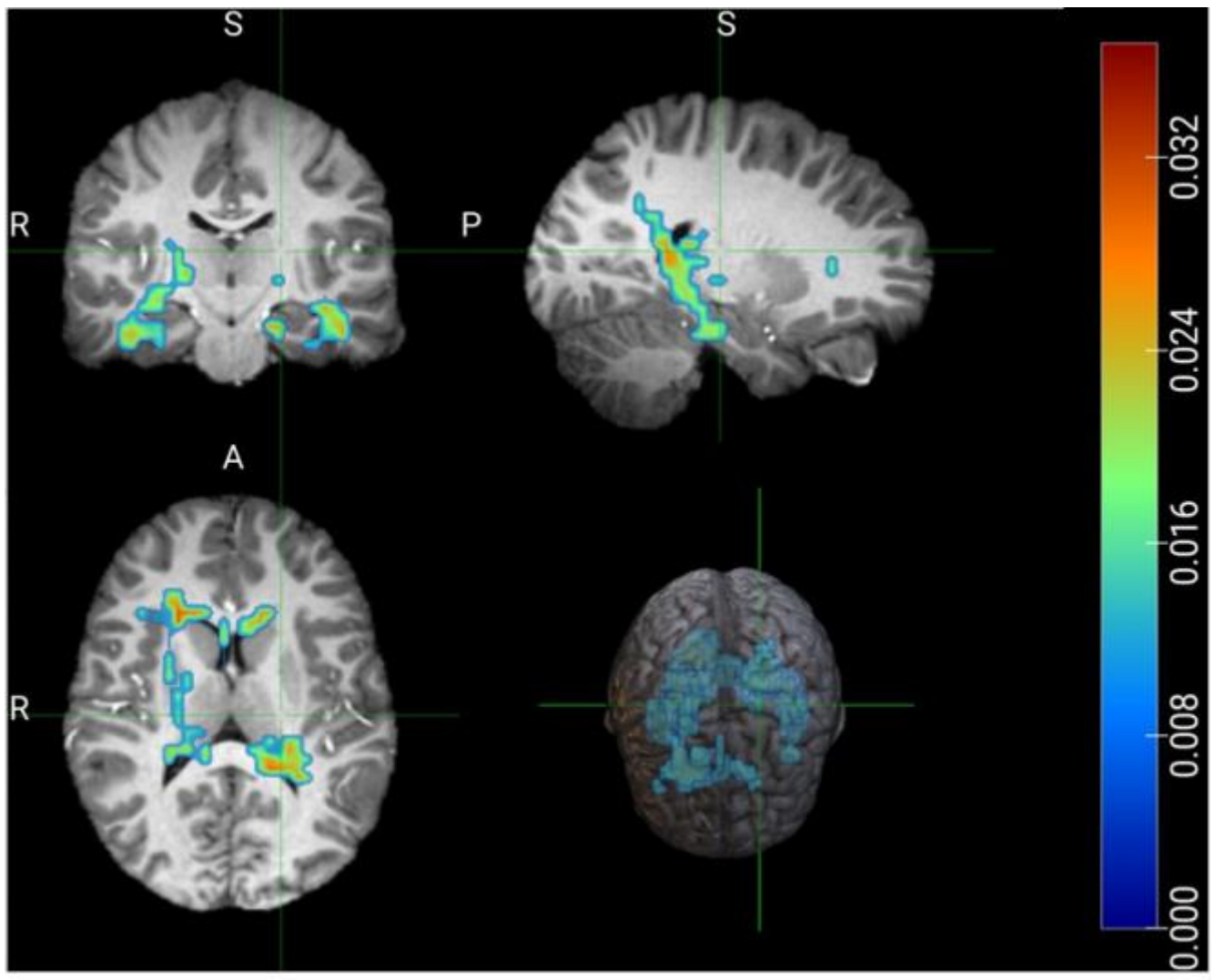

**FIGURE 12** Coefficient map of the estimated $\beta$ for age using Bayesian methods. The results are corrected for multiple comparisons at a significance level of 0.05, and only clusters larger than 50 voxels are retained.

To correct for multiple comparisons within the Bayesian framework, we construct joint credible bands over the entire brain volume using the approach of Ruppert et al. (2003), as applied in similar neuroimaging contexts by Meyer et al. (2015) and Miranda and Morris (2021). This method controls the experimentwise error rate by accounting for the joint distribution of voxelwise estimates. At each MCMC iteration, we calculate the maximum standardized deviation across all voxels to determine a global threshold. Voxels whose $100(1-\zeta)\%$ joint credible intervals exclude zero are flagged as active, ensuring that the probability of any false activation across the brain is bounded by $\zeta = 0.05$ (the significance level of choice here).

To address the sparsity of significant voxels often encountered in whole-brain analyses, after correcting for multiple comparison, we applied a spatial clustering approach. Only spatially contiguous clusters comprising at least 50 voxels were retained for further analysis. This threshold was determined empirically based on the cluster-size distribution observed in our data: using 26-connectivity, the analysis identified 18 clusters ranging from 1 to 810 voxels, with a clear gap separating two large clusters (810 and 52 voxels) from 13 substantially smaller clusters ($\leq$ 15 voxels). The 50-voxel threshold thus retains spatially coherent regions while excluding scattered, isolated activations. This strategy enhances interpretability by focusing on spatially coherent regions of activation rather than isolated voxels. Figure 12 displays the coefficient map for the age variable, highlighting significant regions in the hippocampus and anterior lobe of the cerebellum that are associated with the long-range temporal dependence of the BOLD signal, as modeled by the long-memory parameter.

To assess the robustness of our findings to prior assumptions, we conducted a complementary frequentist analysis using voxelwise linear regression on the projected data. For each voxel, a t-test was performed, and results were corrected for multiple comparisons using the false discovery rate (FDR) at 0.05. Significant voxels were then clustered using the same spatial threshold as in the Bayesian analysis. Figure 13 shows the binary maps for significant voxels for both Bayesian and the frequentist approach and includes a map of their intersection. We observe that both methods identify overlapping regions of significance (highlighted by red circles in panel (c)), providing reassurance that the main spatial patterns detected by the Bayesian analysis are also supported under a frequentist framework. Notably, the Bayesian method via the composite-basis yields bigger regions of activation, reflecting its greater power to detect true effects by accounting for the spatial structure of the data Miranda and Morris (2021).

## 4.3 | Site Effects and Sensitivity Analyses

Prior to harmonization, we examined the extent of site-related heterogeneity in the ADHD-200 dataset. One-way ANOVA revealed significant differences across acquisition sites in Age ($F = 54.49$, $p < 2 \times 10^{-16}$), hippocampal grey matter density ($F = 345.6$, $p < 2 \times 10^{-16}$), and and ADHD severity index ($F = 45.02$, $p < 2 \times 10^{-16}$). As shown in figure 14, age distributions differed substantially across sites, with individual sites covering relatively restricted age ranges. This indicates association

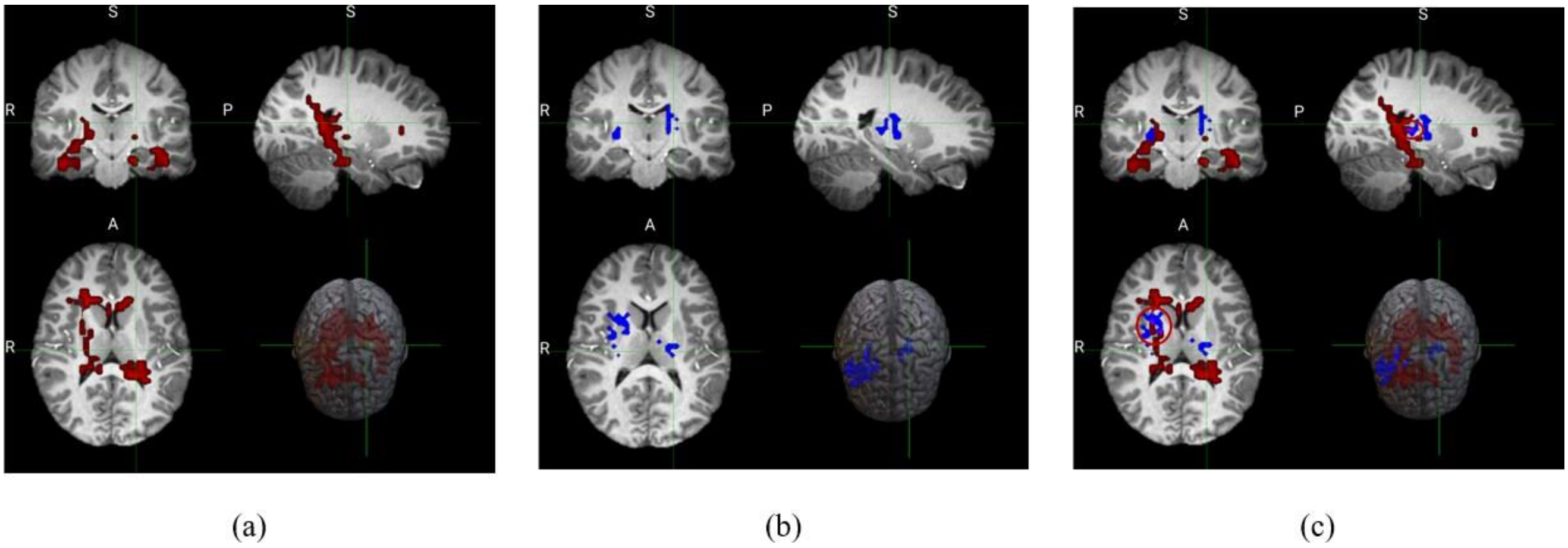

**F I G U R E 13** Binary maps of the estimated $\boldsymbol{\beta}$ coefficients for age using (a) Bayesian and (b) frequentist methods, each corrected for multiple comparisons using an FDR threshold of 0.05 and a cluster-size cutoff of 50 voxels. In (c), yellow regions indicate the intersection of significant voxels between both methods.

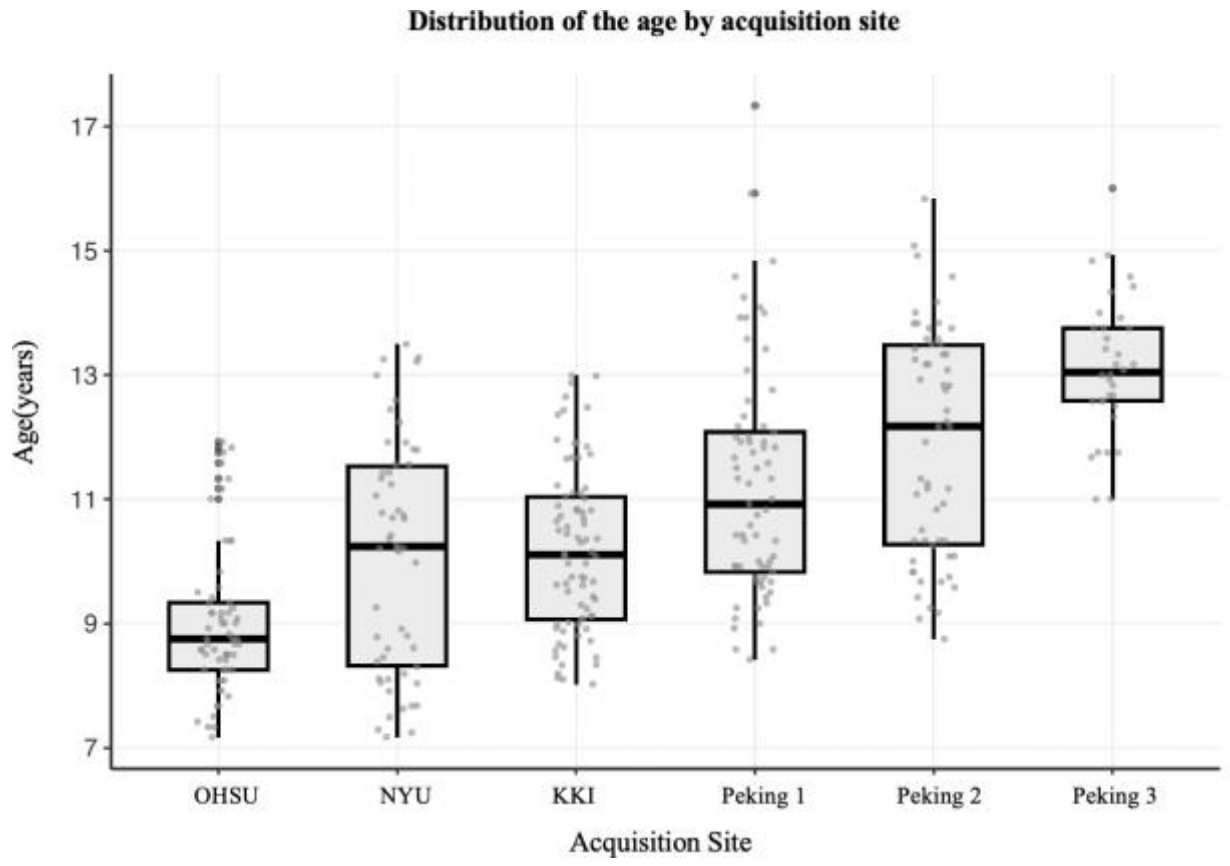

**F I G U R E 14** Distribution of participant age across acquisition sites.The Peking University sample is divided into Peking 1, Peking 2, and Peking 3, representing three acquisition cohorts with differing protocols.

between Age and acquisition site that creates a challenging confounding structure, making it difficult to distinguish age-related variation from scanner and site-related variation in the pooled dataset.

The influence of acquisition group was also evident in the estimated long-memory parameter. the distribution of the estimated LM parameter differed substantially across acquisition groups as shown in figure 15. Together with the ANOVA results, this indicates that acquisition group is associated with both the LM parameter and several biological covariates, creating a challenging confounding structure.

To assess the influence of acquisition group on the Age effect, we refitted the Stage 2 model with acquisition group included as a fixed effect. The mean absolute Age coefficient decreased by approximately 60%, and no voxels remained significant after multiple-comparison correction. However, as shown in Figure 16, the spatial pattern was largely preserved across models, with the strongest effects remaining localized to the same regions and a voxelwise correlation of $r = 0.74$ between the estimated Age coefficient maps. These results indicate that the magnitude and statistical significance of the Age effect are sensitive to acquisition group, whereas its spatial distribution is comparatively stable.

To further assess whether the observed associations reflected within-site age variation, we fitted the Stage 2 model separately within each acquisition site. No significant voxels were identified in any site-specific analysis after applying the multiple-comparison correction at the significance level of 0.05. As shown in Figure 17, the within-site relationships between age and

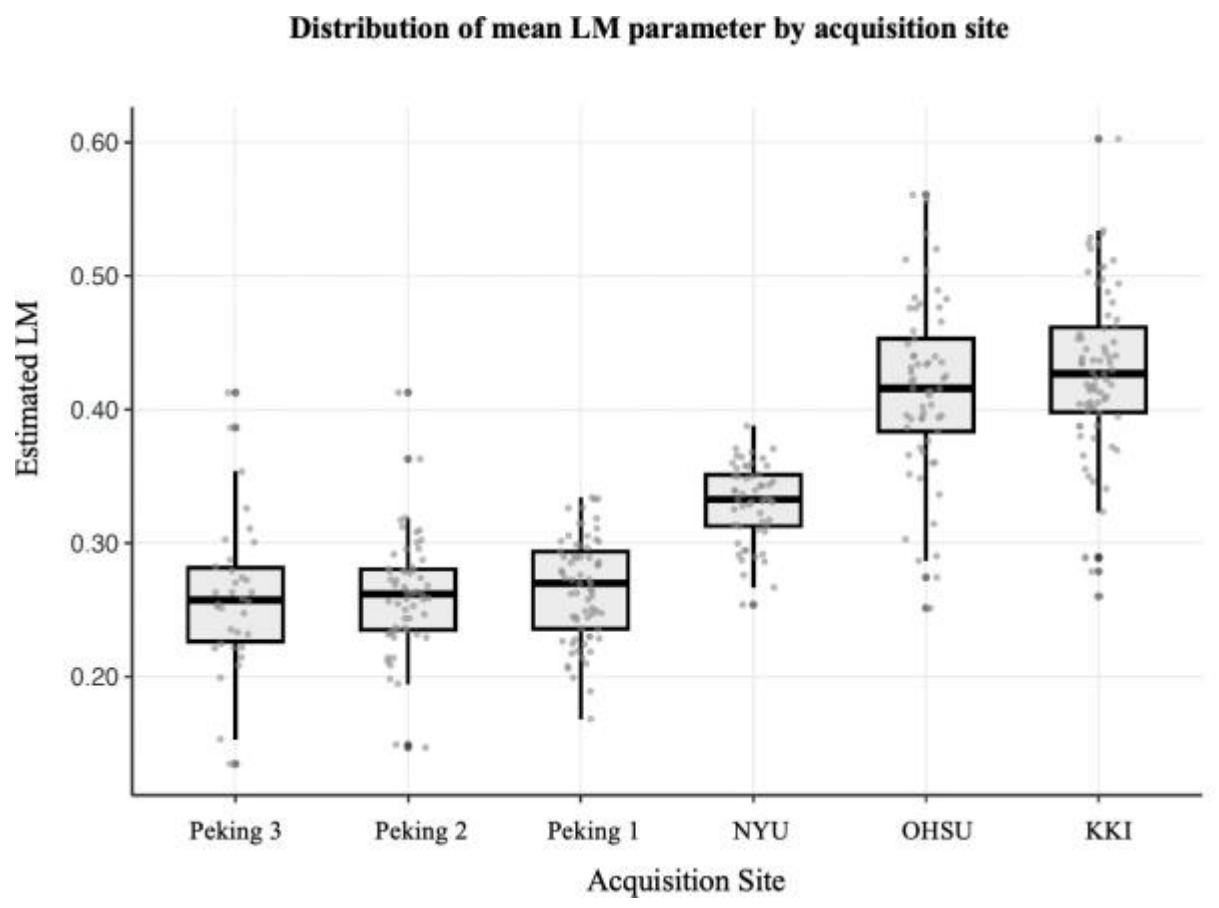

**F I G U R E 15** Distribution of the estimated mean long-memory parameter across acquisition sites. Individual participant estimated LM parameter values are overlaid as jittered points. The number of subjects per site were $n = 78$ for KKI, $n = 54$ for NYU, $n = 61$ for OHSU, and $n = 161$ for Peking.

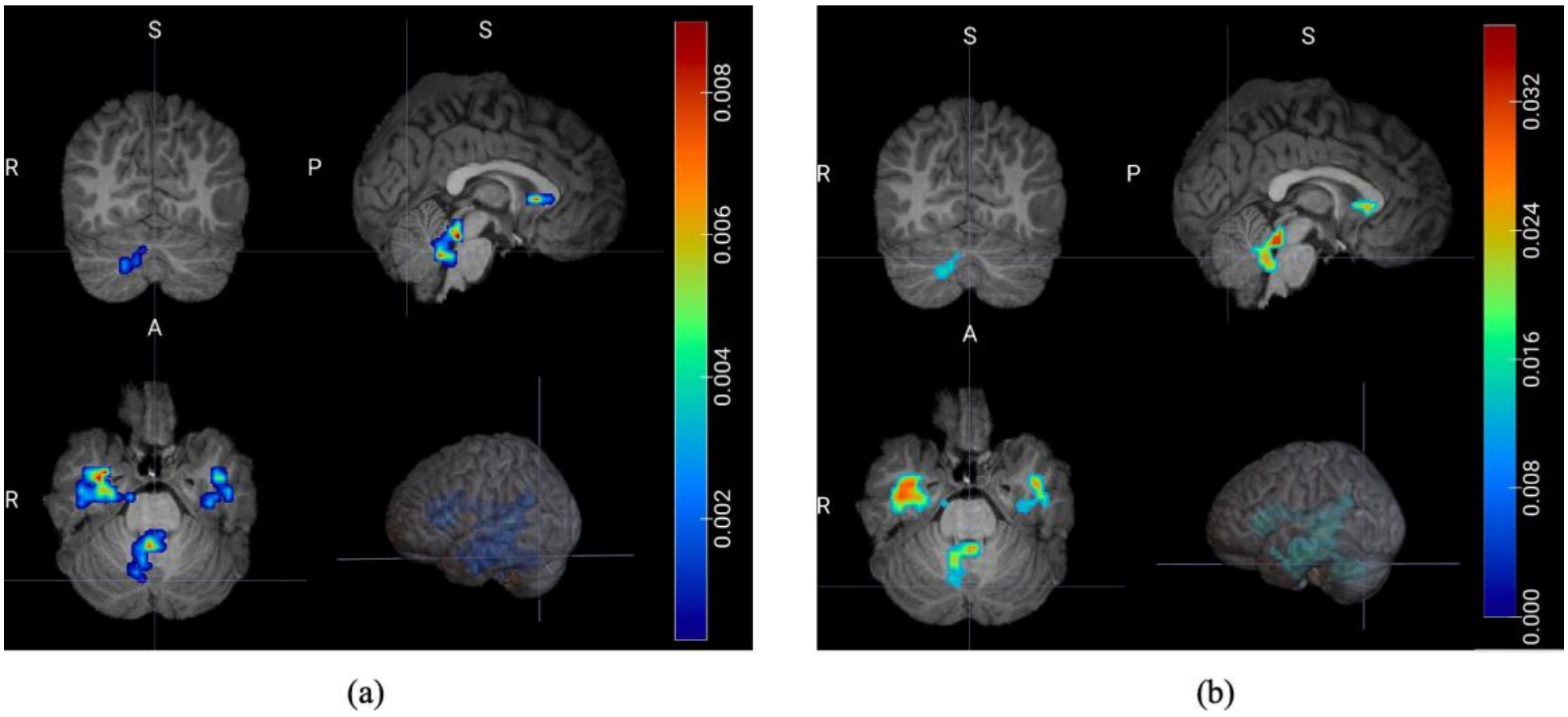

**F I G U R E 16** Coefficient map of estimated $\beta$ for age using the Bayesian Stage 2 model (a) with acquisition site included as a fixed effect in the design matrix; (b) without acquisition site included as a fixed effect in the design matrix.

the mean long-memory parameter were generally weak, with nearly flat trends over restricted age ranges. Together, the limited age variation These results should also be interpreted in the context of the substantially smaller site-specific sample sizes, which ranged from 54 to 161 participants. Together, the limited age variation within sites and the reduced sample sizes decrease the precision of age-effect estimation and substantially reduce statistical power in the voxelwise Bayesian analysis.

Consequently, the strong confounding between Age and acquisition site in the ADHD-200 dataset limits the ability to distinguish within-site age-related variation from scanner-related differences. Consistent with this interpretation, analyses that adjusted for or removed site-related variability also attenuated the evidence for both age and hippocampal grey matter associations (Johnson et al. 2007, Reynolds et al. 2023). These findings indicate that the available data do not provide sufficient information to reliably distinguish biological age effects from acquisition-site variability.

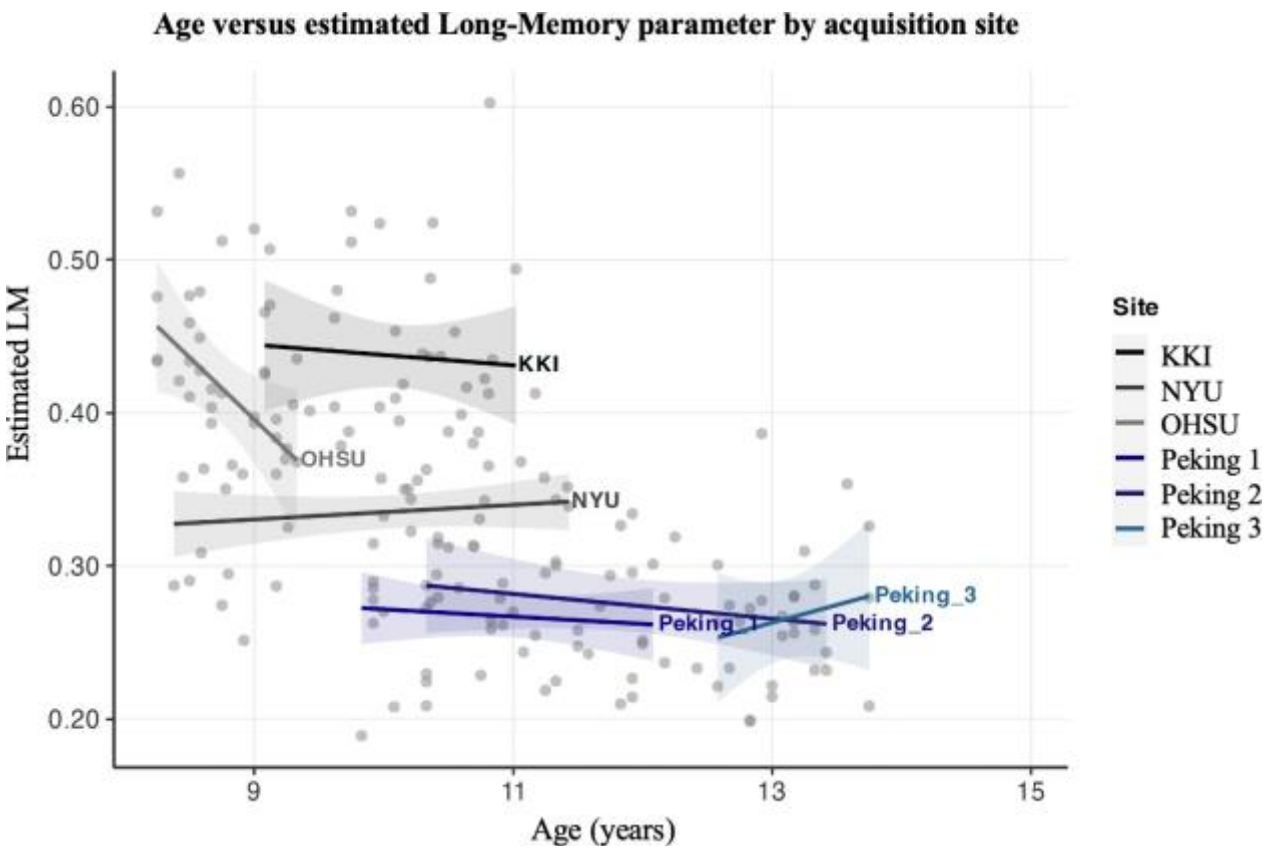

**F I G U R E 17** Relationship between age and the estimated mean long-memory parameter stratified by acquisition site. Each point represents one participant, with colors indicating acquisition site. Solid lines show the site-specific least-squares regression fits, and the shaded regions represent the interquartile age range of participants within each site. The number of subjects per site were $n$ = 78 for KKI, $n$ = 54 for NYU, $n$ = 61 for OHSU, and $n$ = 161 for Peking.

## 5 | DISCUSSION

This study identified a positive association between age and the long-memory properties of resting-state fMRI signals in the hippocampus. The estimated linear regression coefficients indicate that increasing age is associated with higher values of the long-memory parameter, suggesting a developmental trend toward less persistent temporal structure in this region. Notably, no significant associations were observed between the long-memory parameter and the ADHD index, medication status, or their interaction.

Prior studies have investigated the relationship between brain signal complexity and long-memory processes. For example, Sokunbi et al. (2014) used the Hurst exponent to characterize reduced complexity in individuals with schizophrenia, while Park et al. (2018) applied power spectral density (PSD) analysis to differentiate between patient groups, demonstrating its utility as a neuroimaging biomarker. More recently, Medel et al. (2023) demonstrated a robust inverse relationship between the 1/f slope of the PSDa proxy for long-memoryand LempelZiv complexity (LZc), showing that time series with stronger long-range temporal structure (lower LM parameter) exhibited reduced complexity. This relationship was consistently observed across simulations, biophysical models, and empirical recordings. In light of these findings, we interpret the observed decrease in long-memory properties with age (higher $\alpha$ values) as reflecting a developmental shift toward less persistent but more complex brain signal dynamics.

The hippocampus, a medial temporal lobe structure, plays a central role in memory consolidation and spatial navigation. Structural imaging studies have shown that its volume undergoes substantial developmental changes, expanding rapidly in childhood and stabilizing by early adolescence Reitz et al. (2009), Pfluger et al. (1999). Longitudinal studies show that hippocampal volume continues to change during adolescence, with these changes linked to improvements in memory performance Tamnes et al. (2014). Beyond structural growth, functional changes also occur during development. In particular, greater signal complexity has been associated with healthier brain function, reflecting more flexible and efficient neural processing Protzner et al. (2010). Recent work highlights the role of inhibitory circuits in shaping hippocampal plasticity, with evidence that both inhibitory gene expression and long-memory dynamicsindexed by the Hurst exponentstabilize by late childhood Nishio et al. (2024). While hippocampal volume has long served as a marker of cognitive development, limitations of volumetric measures have prompted a shift toward cellular and functional metrics, such as dendritic complexity and neurogenesis Roth et al. (2010). In line with these findings, our study contributes to this literature by showing a positive association between age and the long-memory parameter, $\alpha$, in the hippocampus, suggesting a developmental trajectory characterized by less persistentbut potentially more complexneural signals.

An important consideration in interpreting these findings is the substantial heterogeneity across acquisition sites in the ADHD-200 dataset. Age, hippocampal grey matter density, ADHD Index, and the estimated long-memory parameter differed significantly across sites, demonstrating that acquisition site is strongly confounded with both the response and the biological covariates of interest. When acquisition site was included as a fixed effect in the Stage 2 model, the estimated Age effect was attenuated and no longer reached global statistical significance. However, the spatial pattern was largely preserved, with the strongest estimated effects remaining localized to the same anatomical regions. This suggests that acquisition-site variability has a greater impact on the estimated magnitude and statistical significance of the Age effect than on its anatomical localization. The site-specific analyses provide additional context for this observation. No significant associations were detected when the Stage 2 model was fitted separately within each site, although these analyses were based on substantially smaller sample sizes and much narrower age ranges, both of which considerably reduced statistical power. Consequently, the ADHD-200 dataset provides limited information for estimating age effects independently of acquisition site. This limitation is further supported by the attenuation of the Age effect after adjustment for site-related variability, indicating that the available data do not permit reliable separation of biological age effects from acquisition-site differences.

To the best of our knowledge, this is the first study to analyze voxel-wise temporal complexity of the brain in terms of long-memory properties. We developed a novel rs-fMRI analysis framework that integrates long-memory modeling with Bayesian inference and the discrete wavelet transform (DWT). The DWT offers a multiscale representation of the BOLD signal, which is particularly well-suited to capturing long-range temporal dependencies. This approach is followed by dimensionality reduction via a composite basis and group-level regression analysis to examine associations between spatial patterns of temporal complexity and subject-specific clinical covariates. By applying this framework, we identified associations between long-memory dynamics and age in the hippocampus that are relevant to brain development, offering new insights into the temporal organization of neural activity during childhood and adolescence.

Our approach addresses several key challenges in rs-fMRI analysis, including long-memory temporal dependence, high dimensionality, and the spatial structure inherent in the data. By leveraging a composite basis, we reduce dimensionality while capturing meaningful spatial and temporal patterns. These methodological advances contribute to a deeper understanding of the neural mechanisms involved in brain development and how they are reflected in the temporal dynamics of resting-state brain activity.

A limitation of our two-stage approach is that uncertainty in the first-stage long-memory parameter estimates is not formally propagated to the second-stage group-level inference. To assess the practical impact of this limitation, we conducted a nested MCMC analysis in which Stage 2 regression was performed for each of 800 sampled LM parameter maps from Stage 1 posterior distributions. Comparing the resulting total variance to the posterior varaince from two-stage approach revealed that the missing variance was relatively small across most brain regions (see Supplementary section C Figure C7). While a fully joint hierarchical model would provide more complete uncertainty quantification, the computational costrequiring simultaneous MCMC sampling over 10 million voxel-level parameters, remains prohibitive for whole-brain analysis of this scale. Our two-stage framework offers a pragmatic balance between computational feasibility and statistical rigor, and future work will explore scalable approximate inference methods (e.g., variational Bayes, spatially regularized joint models) to address this limitation more comprehensively.

## 6 | ACKNOWLEDGEMENT

This research was supported by the Canadian Statistical Sciences Institute (CANSSI) through a GSES award, and by the Natural Sciences and Engineering Research Council of Canada (NSERC) through Discovery Grants awarded to Michelle F. Miranda (RGPIN-06941-2020) and Farouk S. Nathoo (RGPIN-04044-2020). We would also like to express our sincere gratitude to Professor Cindy Greenwood for her valuable comments on an earlier version of this manuscript.

# APPENDIX

## A MCMC ALGORITHM

### A.1 First level analysis

To estimate the long-memory parameter for a given subject $i$ and voxel $v$ the joint posterior distribution $(\alpha_{iv}, v_{iv})$ given the wavelet transform timeseries $y^*_v$, we employed a Bayesian hierarchical model. The the joint posterior distribution for a given subject $i$ and voxel $v$ is:

$$
\begin{aligned}
p(v_{iv}, \alpha_{iv}|y^*_{iv}) &\propto (\det(\boldsymbol{\Sigma}_{iv}))^{-1/2} exp(\frac{(y^*_{iv})^T(\boldsymbol{\Sigma}_{iv})^{-1}y^*_{iv}}{-2}) \\
&\times (v_{iv})^{-p-1} exp(\frac{-s}{v_{iv}}) \alpha_{iv}^{a-1}(1-\alpha_{iv})^{b-1}
\end{aligned} \tag{A1}
$$

Under conditional independence across voxels, the subject-level posterior factorizes as $\prod_{v=1}^{N_v} p(v_{iv}, \alpha_{iv} \mid y^*_{iv})$

Where $\Sigma_{iv} = v_{iv} D(\alpha_{iv})$, $\quad D(\alpha_{iv}) = \text{diag}(2^{-m\alpha_{iv}})$.

Having the joint posterior density, the full conditional distribution for $v_v$ is

$$
p(v_{iv}|\alpha_{iv}, y^*_{iv}) \propto v_{iv}^{-p-T^*/2-1} exp\left(\frac{-1}{v_{iv}}\left(\frac{(y^*_{iv})^T\Sigma^{-1}_{iv}y^*_{iv}}{-2}+s\right)\right) \tag{A2}
$$

$$
v_{iv}|\alpha_{iv}, y^*_{iv} \sim inv\text{-}gamma\left(p' = p + \frac{T^*}{2},\ s' = s + \frac{1}{2}(y^*_{iv})^T\Sigma^{-1}_{iv}y^*_{iv}\right) \tag{A3}
$$

and the full conditional distribution for $\alpha$ is

$$
p(\alpha_{iv}|v_{iv}, y^*_{iv}) \propto (\det(\Sigma_{iv}))^{-1/2} \exp\left(-\frac{1}{2}(y^*_{iv})^T\Sigma^{-1}_{iv}y^*_{iv}\right) \alpha_{iv}^{a-1}(1-\alpha_{iv})^{b-1} \tag{A4}
$$

where $y^*_{iv}$ is the transformed fMRI time series signal, $\Sigma_{iv}$ is the covariance function in the wavelet domain.

Figures A1 and A2 present examples of MCMC trace plots, density plots, and autocorrelation function (ACF) plots for the parameters, indicating good convergence.

### A.2 Second level analysis

For the multi-subject analysis, we assumed a Gaussian likelihood for the observed data. The likelihood and prior distributions are specified as follows:

$$
p(\alpha^*|Z, \beta^*, \delta^2) \propto (\delta^2)^{-N/2} exp\frac{(\alpha^* - Z\beta^*)^T(\alpha^* - Z\beta^*)}{-2\delta^2} \tag{A5}
$$

$$
p(\beta^*|\delta^2) \propto |\Lambda_0|^{-Q/2} exp(\frac{(\beta_0 - \mu_0)^T\Lambda_0^{-1}(\beta_0 - \mu_0)}{-2}) \tag{A6}
$$

$$
p(\delta^*) \propto (\delta^2)^{-k-1} exp(\frac{-l}{\delta^2}) \tag{A7}
$$

The combination of these priors and the likelihood function forms the basis of our Bayesian inference, which leads to the full conditionals

$$
\beta^*|\alpha^*, \delta^2 \propto MVN(\mu_n = (\Lambda_n)^{-1}(\mu_0^T\lambda_0 + Z^T\alpha^*), \Lambda_n = (Z^TZ + \Lambda_0^{-1})^{-1}\delta^2) \tag{A8}
$$

$$
\delta^2|\beta^*, \alpha^* \sim Inv - G(k_n = k_0 + q/2, l_n = l_0 + 0.5((\alpha^*)^T\alpha^* + \mu_0\Lambda_0\mu_0 + \mu_n\Lambda_n\mu_n)) \tag{A9}
$$

where $\alpha^*$ is the vector of long-memory parameters estimated in first stage for all subjects, Z is the design matrix, $\beta^*$ is the vector of regression coefficients, and $\delta^2$ is the error variance.

Figure A3 displays example trace plots of the regression coefficients.

## B SIMULATION STUDY

Table B1 shows detailed accuracy measurement calculation of model robustness to variance progression value.

Figure B4 illustrates mean estimated LM maps for the intensity modification scenario, included here as a qualitative visualization of how increasing LM strength in the lower-left region is reflected in the recovered spatial patterns.

Figure B5 illustrates mean estimated LM maps for the radius modification scenario, included as a qualitative visualization of the simulated spatial configurations at different region sizes.

| $v$ Level | Metric | Circle Region LL | LR | UL | UR |
|---|---|---|---|---|---|
| 0.50 | MSE | 0.00267±0.00045 | 0.00259±0.00034 | 0.00269±0.00031 | 0.00282±0.00042 |
| | RMSE | 0.05147±0.00427 | 0.05074±0.00339 | 0.05175±0.00297 | 0.05293±0.00394 |
| | Bias | 0.04090±0.00564 | 0.01844±0.00378 | 0.01473±0.00392 | 0.01204±0.00423 |
| 3.22 | MSE | 0.00250±0.00034 | 0.00245±0.00032 | 0.00271±0.00031 | 0.00259±0.00045 |
| | RMSE | 0.04992±0.00329 | 0.04938±0.00325 | 0.05196±0.00302 | 0.05069±0.00433 |
| | Bias | 0.03986±0.00415 | 0.01275±0.00511 | 0.00871±0.00410 | 0.00487±0.00472 |
| 5.94 | MSE | 0.00236±0.00044 | 0.00245±0.00033 | 0.00246±0.00038 | 0.00254±0.00028 |
| | RMSE | 0.04839±0.00459 | 0.04934±0.00338 | 0.04944±0.00379 | 0.05035±0.00280 |
| | Bias | 0.03749±0.00519 | 0.01264±0.00344 | 0.00812±0.00447 | 0.00325±0.00481 |
| 8.67 | MSE | 0.00237±0.00028 | 0.00230±0.00024 | 0.00269±0.00029 | 0.00273±0.00031 |
| | RMSE | 0.04856±0.00296 | 0.04789±0.00248 | 0.05180±0.00285 | 0.05213±0.00294 |
| | Bias | 0.03679±0.00436 | 0.01297±0.00426 | 0.00624±0.00410 | 0.00569±0.00565 |
| 11.39 | MSE | 0.00240±0.00028 | 0.00231±0.00038 | 0.00261±0.00029 | 0.00264±0.00038 |
| | RMSE | 0.04891±0.00283 | 0.04791±0.00379 | 0.05098±0.00298 | 0.05121±0.00384 |
| | Bias | 0.03834±0.00359 | 0.01170±0.00425 | 0.00694±0.00511 | 0.00425±0.00450 |
| 14.11 | MSE | 0.00243±0.00035 | 0.00237±0.00031 | 0.00259±0.00033 | 0.00285±0.00048 |
| | RMSE | 0.04921±0.00363 | 0.04859±0.00319 | 0.05079±0.00328 | 0.05317±0.00459 |
| | Bias | 0.03783±0.00506 | 0.01105±0.00422 | 0.00696±0.00416 | 0.00550±0.00399 |
| 16.83 | MSE | 0.00238±0.00046 | 0.00256±0.00034 | 0.00266±0.00038 | 0.00263±0.00042 |
| | RMSE | 0.04852±0.00474 | 0.05049±0.00336 | 0.05146±0.00374 | 0.05108±0.00411 |
| | Bias | 0.03712±0.00546 | 0.01170±0.00431 | 0.00794±0.00444 | 0.00401±0.00377 |
| 19.56 | MSE | 0.00252±0.00031 | 0.00233±0.00031 | 0.00258±0.00039 | 0.00250±0.00031 |
| | RMSE | 0.05006±0.00313 | 0.04818±0.00321 | 0.05063±0.00383 | 0.04987±0.00313 |
| | Bias | 0.03895±0.00458 | 0.01143±0.00346 | 0.00728±0.00408 | 0.00343±0.00495 |
| 22.28 | MSE | 0.00238±0.00032 | 0.00238±0.00042 | 0.00258±0.00026 | 0.00264±0.00040 |
| | RMSE | 0.04868±0.00328 | 0.04857±0.00420 | 0.05068±0.00255 | 0.05125±0.00399 |
| | Bias | 0.03782±0.00449 | 0.01207±0.00432 | 0.00679±0.00476 | 0.00490±0.00473 |
| 25.00 | MSE | 0.00226±0.00030 | 0.00245±0.00036 | 0.00272±0.00037 | 0.00269±0.00035 |
| | RMSE | 0.04743±0.00335 | 0.04933±0.00363 | 0.05208±0.00351 | 0.05178±0.00342 |
| | Bias | 0.03697±0.00469 | 0.01339±0.00493 | 0.00719±0.00415 | 0.00329±0.00402 |

**TABLE B1** Region-specific estimation error (mean±SD across 25 subjects) for each activated region under varying innovation variance $v$.

## C VARIANCE STUDY

A referee has noted some concern about whether parameter variability from the first stage is properly propagated to the second stage. To assess the impact of not propagating Stage 1 uncertainty in our two-stage framework, we conducted a nested MCMC analysis following the law of total variance. Specifically, we decompose the total variance of regression coefficients into within-stage and between-stage components:

$$\mathrm{Var}(\beta|Y) = \underbrace{E_{\alpha}[\mathrm{Var}(\beta|\alpha, Y)]}_{\text{Term A: within-stage}} + \underbrace{\mathrm{Var}_{\alpha}[E(\beta|\alpha, Y)]}_{\text{Term B: propagated uncertainty}} \tag{C10}$$

**Nested MCMC Implementation:** We drew 800 thinned samples $\alpha_i^{(j)}$ ($j = 1, ..., 800$) from the Stage 1 posterior distributions for each subject $i$ across all voxels $v$. For each $\alpha_i^{(j)}$, we performed a complete Stage 2 analysis (PCA projection followed by Bayesian regression), generating 1,000 posterior samples $\beta_v^{(j,s)}$ ($s = 1, ..., 1,000$) for each voxel $v$. The variance components were computed as:

**Within-stage variance (Term A):** Average conditional variance across $\alpha$ samples:

$$\bar{v}_v = \frac{1}{800} \sum_{j=1}^{800} \mathrm{Var}_s(\beta_v^{(j,s)}) \tag{C11}$$

**Propagated variance (Term B):** Variance of conditional means across $\alpha$ samples:

$$v_{\text{prop},v} = \text{Var}_j\, \bar{\beta}_v^{(j)}, \quad \text{where} \quad \bar{\beta}_v^{(j)} = \frac{1}{1000}\sum_{s=1}^{1000} \beta_v^{(j,s)} \tag{C12}$$

**Total variance:**

$$v_{\text{total},v} = \bar{v}_v + v_{\text{prop},v} \tag{C13}$$

**Two-Stage Approach:** In our standard two-stage analysis, we computed posterior means $\hat{\alpha}_{iv} = \text{E}[\alpha_{iv}|y^*_{iv}]$ from Stage 1 for each subject $i$ and voxel $v$, then performed Stage 2 regression once treating $\hat{\alpha}_i = \{\hat{\alpha}_{iv}, v = 1, ..., N_v\}$ as fixed. The posterior variance from this single Stage 2 MCMC run:

$$v_{\text{two-stage},v} = \text{Var}(\beta_v|\hat{\alpha}, Y) \tag{C14}$$

This captures only Term A (within-stage uncertainty), as the uncertainty in $\alpha_{iv}$ is not propagated to the group-level inference. Figure C7 compares the total variance (panel a) with the two-stage variance (panel b) for the age covariate. The spatial patterns show that while some uncertainty is lost by not propagating Stage 1 parameter uncertainty, the two-stage approach captures the total variance across most brain regions to some good extent. The missing variance ($v_{\text{total}} - v_{\text{two-stage}}$) is relatively modest, suggesting that our two-stage inference remains well-calibrated for detecting group-level associations.

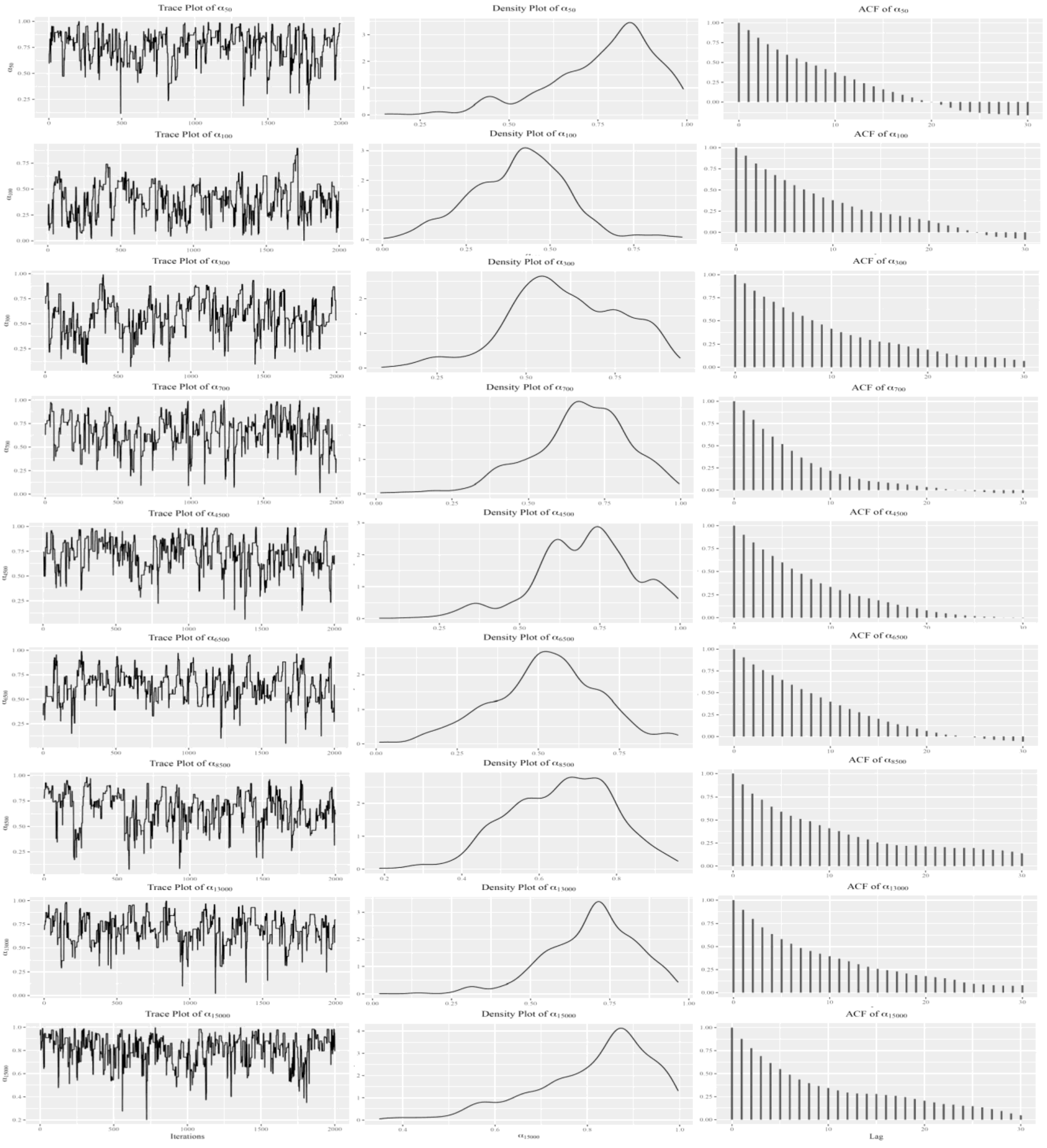

**FIGURE A1** Diagnostic plots for the long-memory parameter, $\alpha$, at randomly selected voxels. (a) Trace plot. (b) Density plot. (c) Autocorrelation function (ACF) plot.

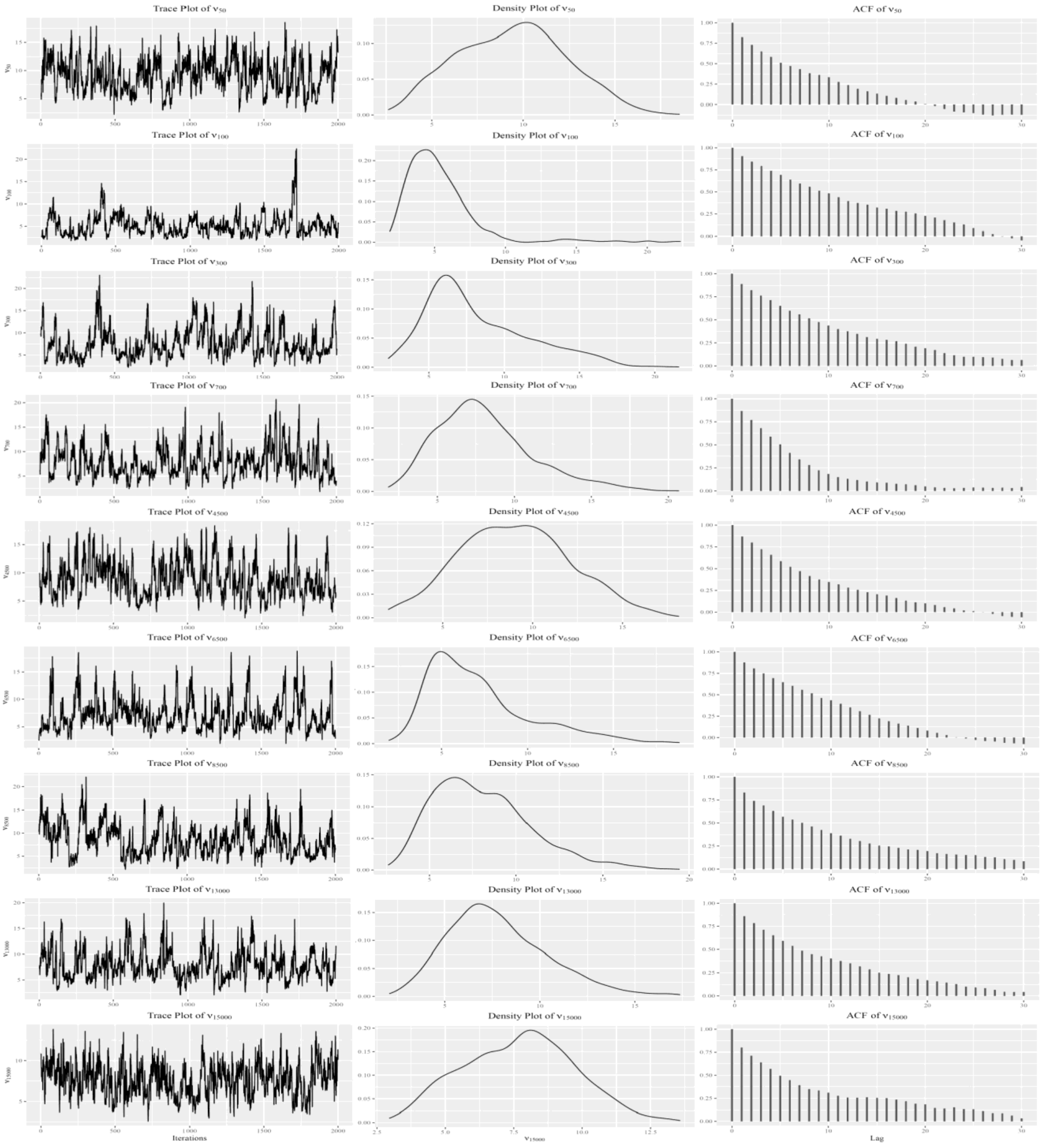

**FIGURE A2** Diagnostic plots for the innovation variance, *v*, parameter at randomly selected voxels. (a) Trace plot. (b) Density plot. (c) Autocorrelation function (ACF) plot.

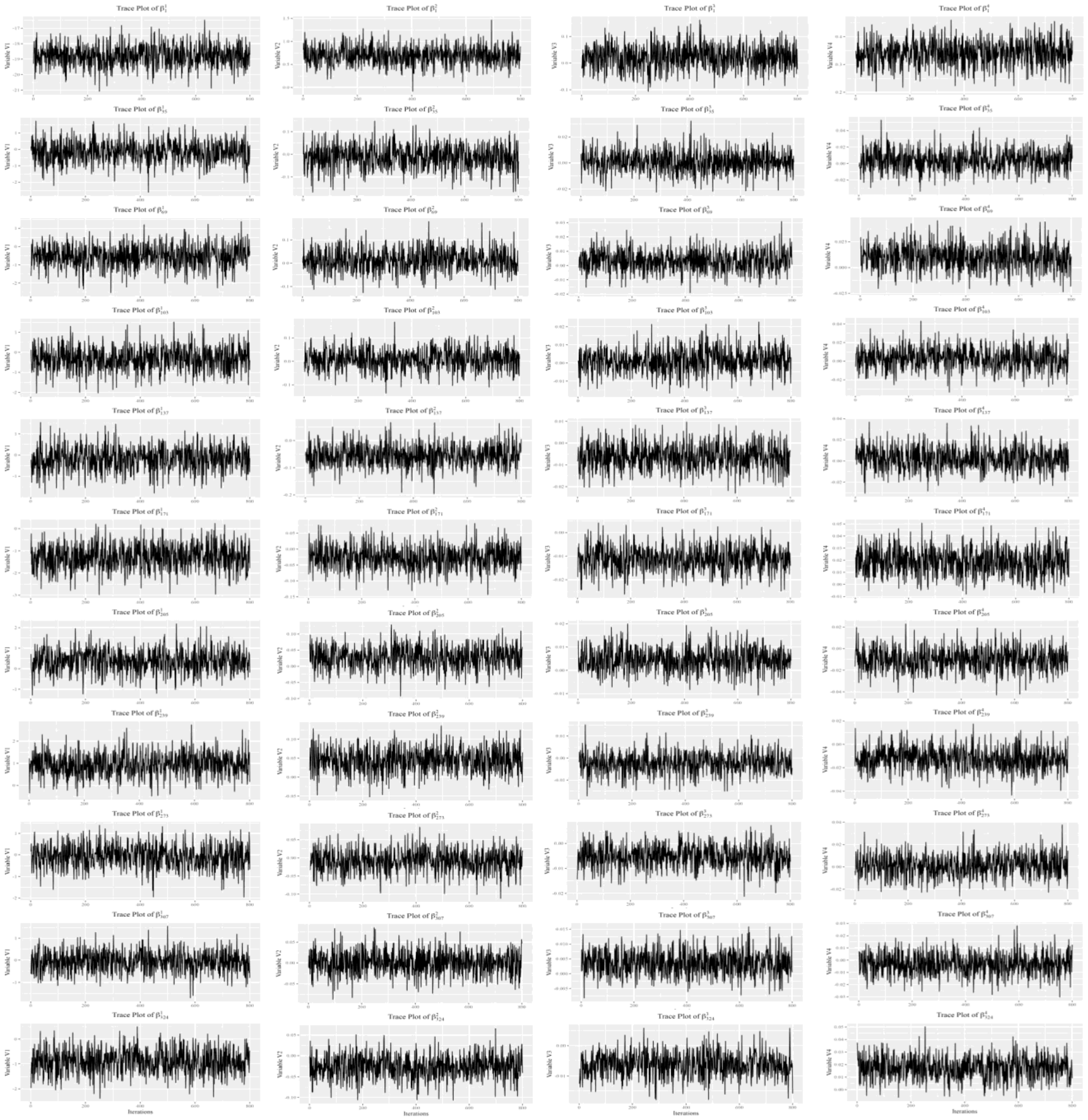

**F I G U R E A3** Trace plots of covariates projected onto principal component (PC) space. (a) Age. (b) Medication status. (c) ADHD index. (d) Interaction between ADHD index and medication status.

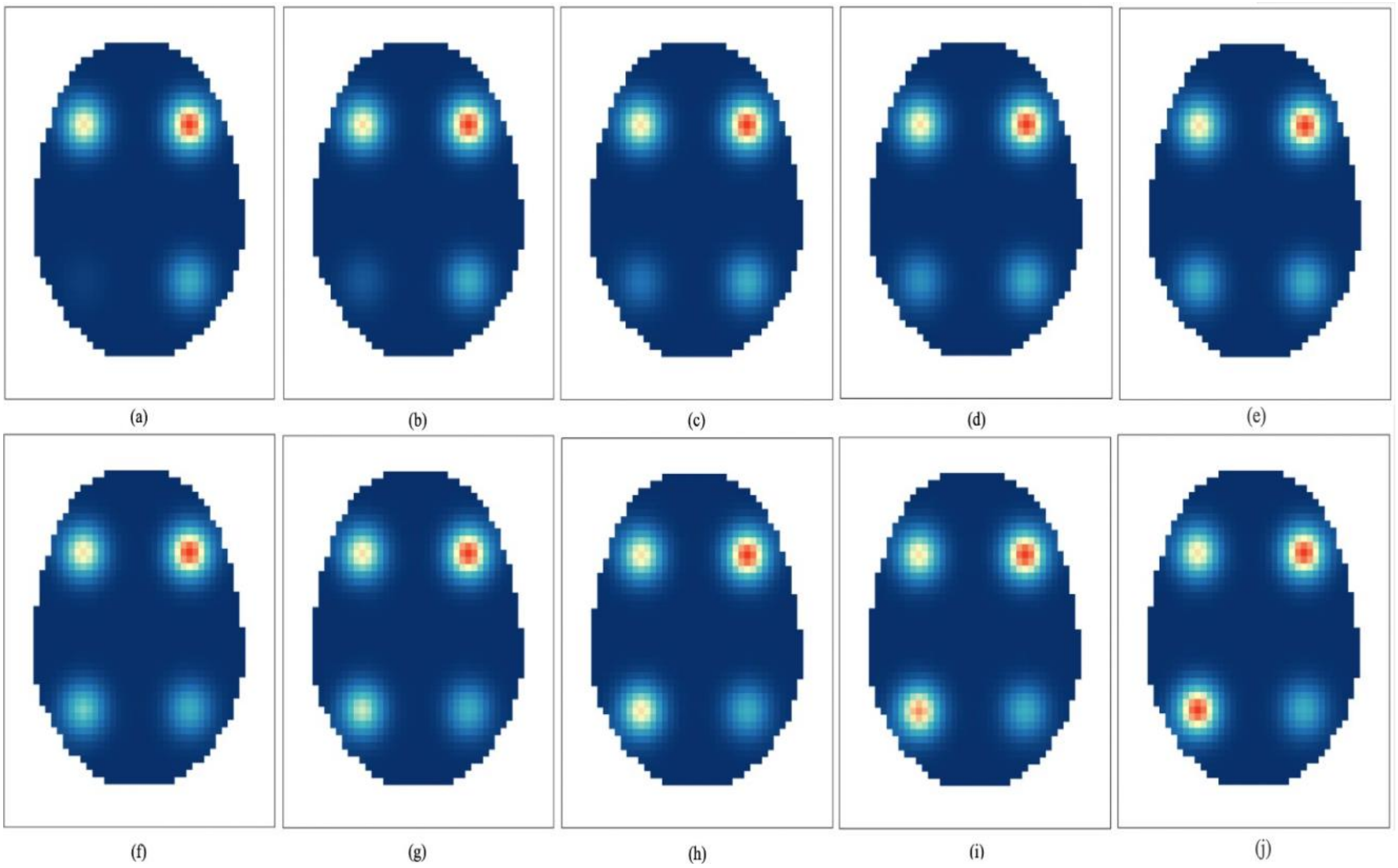

**F I G U R E B4** Mean estimated LM maps under the intensity modification scenario. Panels (a)–(j) correspond to increasing LM levels in the lower-left region, ranging from 0.05 to 0.95 in increments of 0.1, while all other regions remain fixed. Each panel shows the mean estimated LM map averaged across simulated subjects.

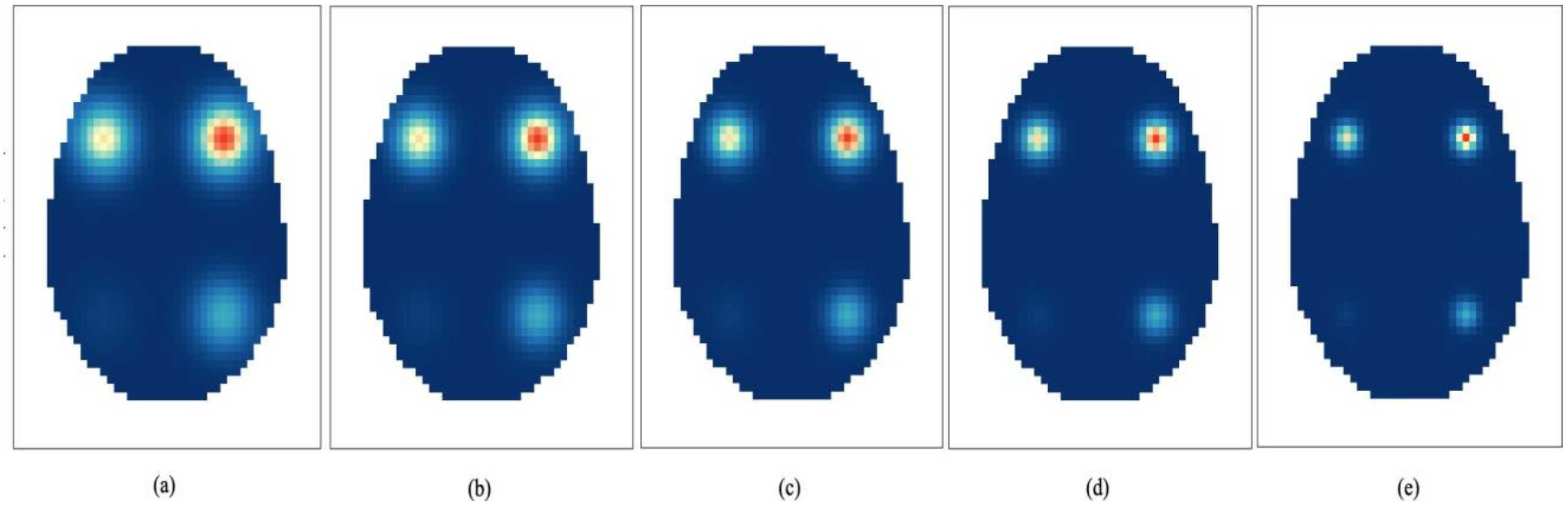

**F I G U R E B5** Mean estimated LM maps under the radius modification scenario. Panels (a)–(e) correspond to decreasing activation radius levels, with region size progressively reduced while all other simulation components are held fixed. Each panel shows the estimated LM map averaged across simulated subjects, using our stage one proposed model.

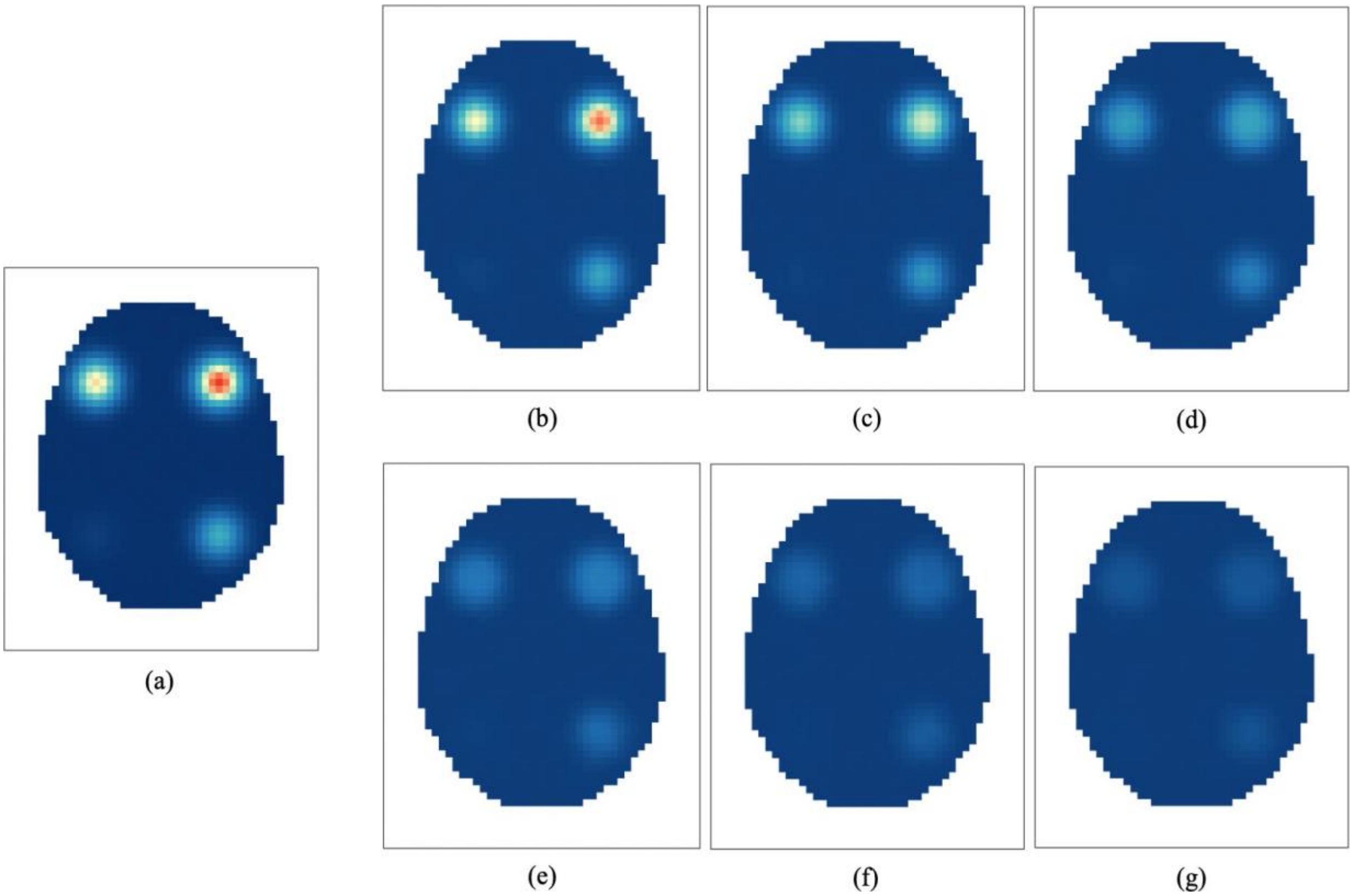

**F I G U R E B6** Noise perturbation scenario. Panel (a) shows the true LM map. Panels (b)–(g) display mean estimated LM maps averaged across subjects under increasing additive noise levels $C \in \{0.2, 0.4, 0.6, 0.8, 1.0\}$, illustrating progressive attenuation of spatial structure as noise dominates the signal.

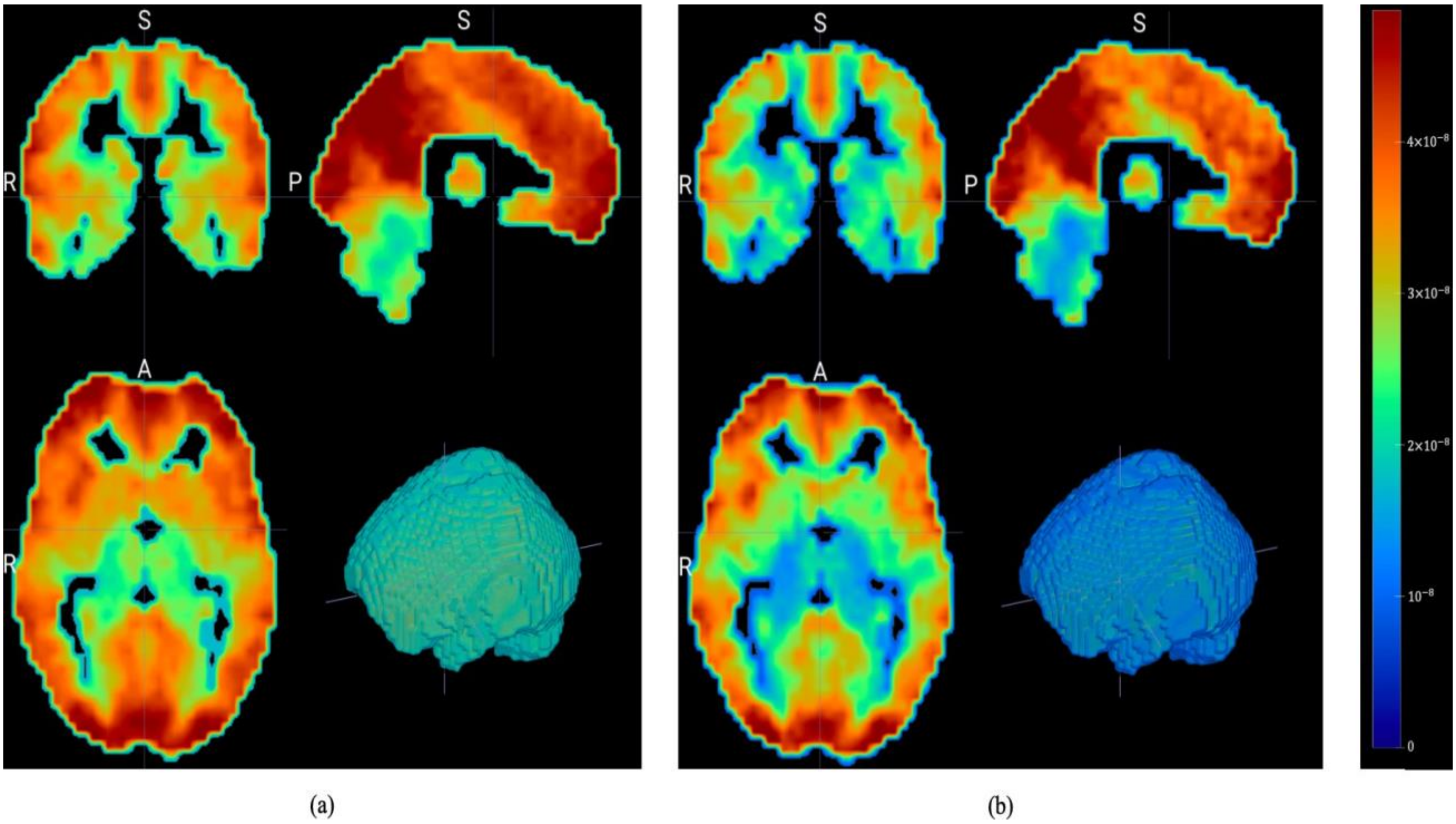

**F I G U R E C7** Comparison of variance estimates from nested MCMC and two-stage approaches for the age covariate;(a) Total variance map obtained from nested MCMC analysis (800 outer iterations Œ 1,000 inner MCMC samples),(b) variance map from the two-stage approach.